# System Effects of Carbon-Free Electricity Procurement: Regional Technology and Emissions Impacts of Voluntary Markets

John Bistline[1]*, Geoffrey Blanford[1], Adam Diamant[1], Arin Kaye[1], Daniel Livengood[1], Qianru Zhu[1], Francisco Ralston Fonseca[1]

**Affiliations:**

[1] Electric Power Research Institute; Palo Alto, USA.

*Corresponding author. Email: jbistline@epri.com.

## Abstract

Voluntary carbon-free electricity (CFE) procurement has the potential to accelerate electric sector decarbonization, but procurement strategies vary widely, leading to uncertainty about emissions, investments, and costs. This study assesses the system-wide effects of voluntary CFE procurement on U.S. regional power systems using a detailed energy systems model across a range of program designs, eligible technologies, policy environments, and modeling assumptions. Results suggest that hourly matching—where clean electricity procurement aligns with hourly load—combined with new and local generation could maximize emissions reductions from CFE procurement, particularly under existing Inflation Reduction Act incentives and state policies. However, regional costs vary significantly, with a CFE cost premium ranging from $11-63/MWh nationally across scenarios and $1-130/MWh across regions, broader than previous estimates. Expanding the eligible technology portfolio to include renewables, nuclear, carbon capture, and energy storage reduces costs, particularly in regions with lower wind and solar resource quality, though variable renewables and battery storage remain the dominant resources in many scenarios. Additionally, we show that the future policy environment strongly influences the effectiveness of voluntary CFE programs, with more stringent emissions policies or subsidies potentially limiting the incremental benefits of procurement. The analysis also quantifies how features of the model framework can shape insights about CFE procurement strategies.

**Introduction**

There is growing interest in carbon-free electricity procurement by electric companies, large electricity customers, and other stakeholders, including procuring and supplying carbon-free electricity 24 hours per day, 7 days per week (24/7 CFE). This trend represents an evolution of corporate clean energy procurement to match procurement with real-time hourly load, which complements policies to drive emissions reductions [1, 2]. This market is small but growing due to large customers (including companies such as Google and Microsoft with growing data center loads [3]), hydrogen tax credits through the Inflation Reduction Act, and Executive Order 14057, which required the U.S. Federal Government to procure 100% CFE by 2030 on an annual basis, including 50% hourly matched CFE [4]. This trend toward hourly CFE matching (where clean electricity procurement aligns with load on an hourly basis) differs from the more common practice of annual matching (where consumer demand and CFE procurement are aligned on a volumetric basis over a year).

Another motivation is the revision of the Greenhouse Gas Protocol for corporate emissions accounting, which is expected to be finalized in 2025 [5]. Current guidance allows matching of annual electricity consumption on a volumetric basis. However, there is growing interest in hourly matched CFE, which may be facilitated by hourly Energy Attribute Certificates (EACs), which are tradeable instruments representing generation from eligible resources that can be either bundled or unbundled from the underlying energy.

Previous studies quantified how annual (volumetric) matching can lead to more limited $CO_2$ reductions than anticipated [6, 7, 8]. However, research on system-level implications of CFE demand has focused analysis on a limited number of regions. For instance, Xu, et al. (2024) conduct analysis for two regions in the Western U.S., which have good wind and solar resource quality and stringent state-level policies that could make results less generalizable. Existing studies also focus on temporal matching under current policies, rather than across other assumptions about program design and policy conditions. Studies on Inflation Reduction Act (IRA) clean hydrogen 45V tax credits [9, 10, 11] and EU electrolytic hydrogen regulations [12, 13] quantify impacts of an important subset of CFE demand under "three pillar" qualification criteria—temporal matching (i.e., hourly CFE generation must coincide with hourly consumption), incrementality (i.e., CFE resources must be new capacity), and deliverability (i.e., CFE generation must reside in the same region as demand). However, insights from these studies may not apply to other sources of electricity demand, given electrolytic hydrogen's unique load shapes and design of these incentives.

This analysis assesses potential system effects of voluntary CFE procurement across a range of assumptions about program design and participation, eligible technologies, policy environment, and modeling framework as well as the regional variation in these impacts. Here we extend the existing literature by quantifying:

- Effects of CFE procurement under a wider range of assumptions about program design and the policy environment, including regional impacts. This regional analysis considers a wider range of grid settings than earlier studies (16 regions, instead of 2-4 [6, 8, 14]), illustrating the variation in CFE procurement impacts across different policies and resource mixes, which gives estimates for the cost and emissions effects of hourly matching. The analysis also provides the first analysis on the impacts of the size of deliverability regions.

- Impacts of qualifying technologies and the value of broader technological portfolios. This work builds on earlier analysis [8, 15] to illustrate the cost impacts of technology availability and differences across regions.
- Effects of model choices. Given the analytical challenges associated with representing CFE procurement, we quantify how key decisions about the modeling framework could alter insights about CFE procurement costs and technology strategies. We are the first to assess the impacts of alternate weather years, temporal resolution, and detailed end-use load shape modeling. This study uses an integrated energy systems model to study these questions, unlike earlier CFE procurement studies that use models of the power sector only [6, 8].
- Interactions between voluntary CFE demand and other claims on clean electricity from state-level policies and electrolytic hydrogen credits. We also conduct sensitivities to alternate future climate policy environments, which builds on earlier studies that focus on current policy settings.

These findings can inform electric companies planning for decarbonized supply, customers interested in 24/7 CFE, designers of procurement protocols, as well as stakeholders aiming to understand potential impacts of these trends.

**Results**

*Modeling Carbon-Free Electricity Procurement*

This analysis uses EPRI's U.S. Regional Economy, Greenhouse Gas, and Energy (REGEN) model to assess system impacts of CFE procurement across scenarios. REGEN is an integrated energy systems model with a detailed electric sector model that accounts for investments and operations over time for generation, transmission, energy storage, carbon removal, and fuels supply (e.g., electrolytic hydrogen production). The large-scale optimization determines the least-cost mix of resources given assumptions about technology costs, markets, and policies while capturing temporal detail between load, wind output, and solar output as well as chronological operations to characterize energy storage and other balancing resources [16, 17]. This analysis uses full hourly temporal resolution for investment and operational decisions and conducts sensitivities to understand how less temporal resolution could alter insights. The REGEN end-use model provides hourly estimates of regional load over time based on detailed modeling of technology adoption and utilization across buildings, transport, and industrial sectors [18]. More details about REGEN can be found in Methods and detailed documentation [19].

Scenarios for the analysis are summarized in Table 1, which vary CFE program design and participation rate (i.e., market size), qualifying technologies, assumed policy environment, and assumptions about the modeling framework. We run combinations of many of these scenarios to examine interactions, especially of hourly temporal matching. CFE targets are applied as share of commercial and industrial (C&I) load to all model regions simultaneously (as discussed in Methods, these segments of demand total 66% nationally by 2035); program participation is varied between 10% and 50% of C&I load with a focus on outcomes in 2035.

**Table 1. Summary of scenario configurations and abbreviations.** Detailed descriptions are provided in the Methods section and Supplementary Information (Note S2). Default values are shown for each class of sensitivity. Combinations of different configurations are conducted for scenarios in this analysis.

| Configuration (Abbr.) | Description |
|---|---|
| *Carbon-Free Electricity (CFE) Qualification Criteria* | |
| **Reference (ref)** | On-the-books federal and state electric sector policies and incentives, including the Inflation Reduction Act (IRA), but without voluntary CFE demand; no explicit national $CO_2$ policy |
| **Three Pillars (cfe_3p)** | Qualified generation must be zero-emitting and satisfy temporal matching (i.e., hourly CFE generation must coincide with hourly consumption), incrementality (i.e., CFE resources must be new capacity), and deliverability (i.e., CFE generation must reside in the same region as demand) |
| **Temporal Flexibility (cfe_ann)** | Annual/volumetric matching instead of hourly matching but assuming incrementality and deliverability |
| **Resource Flexibility (cfe_ex)** | Existing resources allowed instead of excluded but assuming hourly matching and deliverability |
| **Locational Flexibility (cfe_usa)** | CFE anywhere in the U.S. qualifies instead of only in the 16 model regions in Figure S1 (we also consider sensitivities with intermediately sized regions) but assuming hourly matching and incrementality |
| *Alternative CFE Participation Rate (i.e., Market Size)* | |
| **10% Participation (Default)** | 10% of commercial and industrial electricity demand in 2035 |
| **50% Participation** | 50% of commercial and industrial electricity demand in 2035 |
| *Qualifying Technologies* | |
| **Default** | All zero-emitting technologies, including variables renewables (e.g., wind and solar power), hydro, biomass, geothermal, nuclear, and energy storage |
| **Variable Renewables Energy Only (vre)** | Wind, solar, and batteries only |
| **All Options (all)** | All lower-emitting options, including carbon capture, where carbon dioxide removal (CDR) can be used to offset residual emissions |
| *Alternative Weather Years* | |
| **Default** | 2015 meteorology and temperatures used for hourly time-series variables (e.g., potential wind and solar output) |
| **1999-2019** | Sensitivities consider investments and operations optimized to single weather years from 1999 through 2019 |
| *Model Temporal Resolution* | |
| **Static (Default)** | Hourly model with 8,760 segments for investment and system operations for single future year (2035) |
| **Dynamic** | Intertemporal optimization in five-year periods through 2050 with 120 intra-annual periods and reduced-form chronology |
| *CFE Load Shapes* | |
| **Variable (Default)** | Hourly load profiles are based on outputs from REGEN's end-use model |
| **Flat** | Flat hourly load shapes that match aggregate annual CFE demand |
| *Assumed Policy Environment* | |
| **No IRA (noira)** | Counterfactual without IRA tax credits but all other state-level policies from the reference scenario; all cases assume 10% participation rate |
| **IRA (Default)** | Includes IRA tax credits for the power sector, hydrogen, carbon capture, and end-use electrification |
| **Carbon Fee (cfee)** | Carbon fee starting at $20/t-$CO_2$ in 2025 and rising at 3% annually in real U.S. dollar terms (i.e., above inflation) |

*National Impacts of CFE Procurement*

Model results suggest that relaxing any of three pillars—temporal matching, incrementality, and deliverability—leads to large differences between consequential and attributed generation (Figure 1A), even with the other two pillars in place. Generation providing zero-emission credits is referred to as "attributed" (i.e., resources providing credits to meet the CFE procurement constraint); however, because of responses in power system dispatch and capacity, the actual change in generation differs, which is known as "consequential generation." Consequential changes represent the net effect of adding CFE procurement and require detailed modeling to assess, which is the difference between a scenario with CFE procurement and counterfactual reference without procurement.

Annual matching has more limited consequential effects on system emissions and generation than hourly matching (Figure 1A)—1 million tons (Mt) of $CO_2$ per year lower than the reference with annual matching compared with 42 Mt-$CO_2$/yr with hourly under the 10% participation scenarios. There is inframarginal wind and solar generation with annual matching due to IRA incentives and technological change, which leads to an EAC oversupply and limited incremental low-emitting generation. There are similarly limited $CO_2$ impacts without qualification criteria for new resources or local deliverability.

For scenarios with 50% participation, higher CFE demand leads to smaller deviations from relaxing qualification criteria, since the volume of EAC oversupply is limited relative to CFE demand. Higher CFE participation leads to lower emissions leakage from relaxed qualification (Figure S8): 4-59% of three-pillar $CO_2$ reductions with 50% participation (compared to 1-2% at a 10% participation). Greater participation alters marginal system responses and displaces more coal (relative to gas) due to CFE procurement and increases in wind, solar, and nuclear power generation, especially with three-pillar requirements. These scenarios illustrate the large magnitude of CFE participation that is needed for consequential impacts to approach attributed ones.

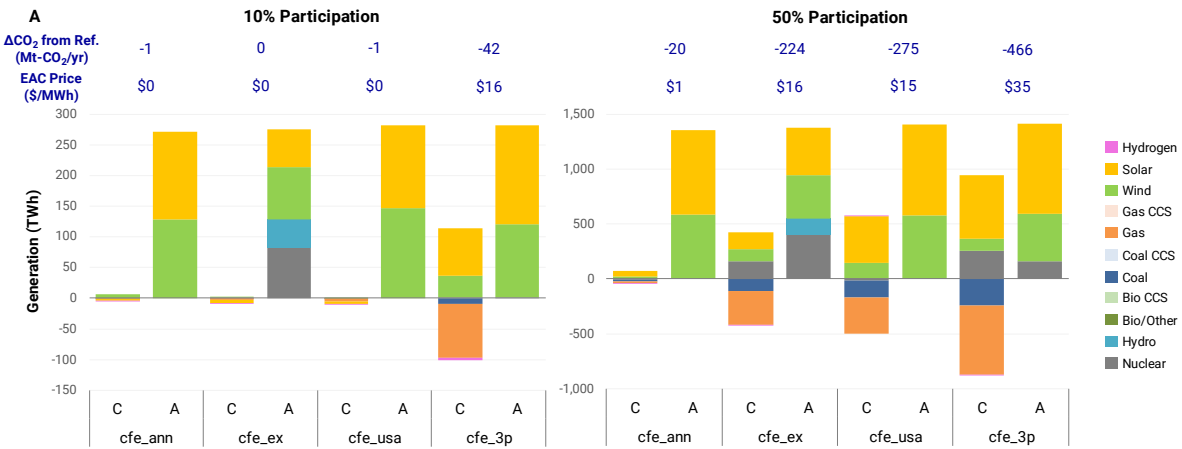

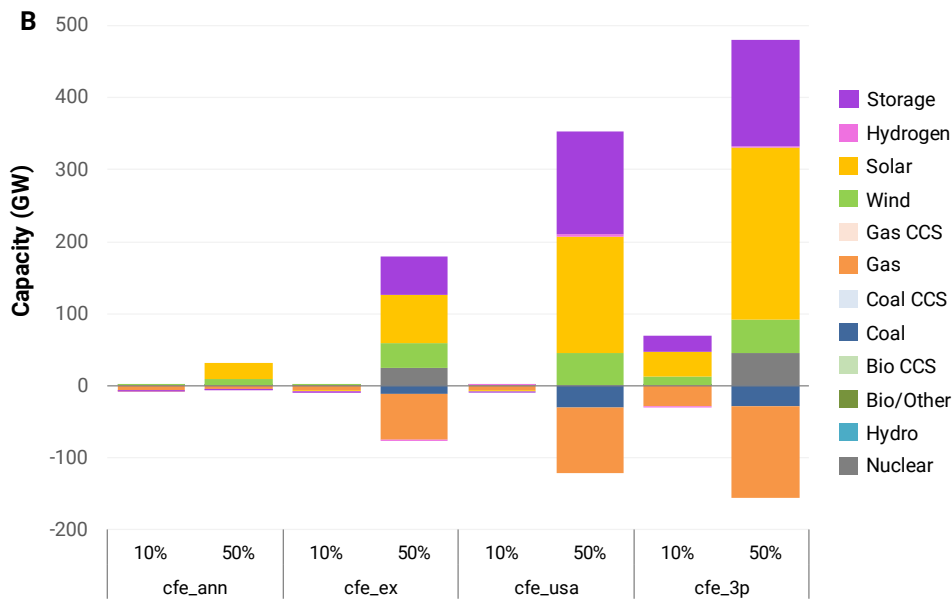

**Figure 1. Changes in national U.S. generation and installed capacity from CFE procurement by scenario in 2035 (relative to the reference without CFE demand).** (A) Changes with 10% and 50% C&I CFE demand. Bars are shown for consequential generation (i.e., actual change in generation) and attributed generation (i.e., providing energy attribute certificates for qualification). (B) Consequential changes in installed capacity across scenarios. C = consequential impacts; A = attributed impacts; CCS = carbon capture and sequestration. The energy attribute credit (EAC) price is the generation-weighted average of shadow prices on the CFE procurement constraint across regions and hours. Changes in emissions are shown relative to the no CFE reference scenario, which are 960 Mt-$CO_2$/yr in 2035.

Figure 1B shows changes in installed capacity across scenarios. Hourly matching leads to a large expansion of energy storage with up to 149 GW increase with 50% participation relative to the reference. As discussed in the next section, energy storage deployment varies regionally and is primarily battery storage with increasing durations for deeper decarbonization. In contrast, there is minimal incremental energy storage procurement with annual matching.

The combination of hourly matching, new resources, and local deliverability can maximize emissions reductions from CFE procurement but at a cost premium that varies by participation level (Figure 1A). The EAC price with three pillars is $16/MWh and $35/MWh nationally for the 10% and 50% participation scenarios, respectively, which is the generation-weighted average of shadow prices on the EAC constraint. Low EAC prices with flexible criteria reflect non-additional procurement (i.e., resources that would have been built by other firms without voluntary purchases).

*Regional Variation in CFE Impacts*

The costs of meeting CFE demand vary by region (Figure 2A). Regional EAC prices are marginal costs associated with meeting procurement requirements above electricity generation prices. Costs of meeting three-pillar qualification criteria are highest in regions with lower wind and solar resource quality (especially in the U.S. East and South, as shown in Figure S2) and lowest in regions with better resources (especially in the Midwest). Note that some regions with binding emissions policies or mandates (e.g., for offshore wind) have lower EAC prices due to the higher amount of qualified CFE they bring in the baseline (Figure 4). Three-pillar EAC prices increase with participation rate—$1-36/MWh with 10% rate ($16/MWh average across the U.S.) versus $20-54/MWh with 50% rate ($35/MWh average nationally). These values are higher than the price with national deliverability of $15/MWh for 50% participation (Figure 1A), where regional trading leads to a single national price.

These regional differences raise questions about which regions are best suited for implementing hourly matching, especially if there is locational flexibility in siting loads. In general, there is a tradeoff between the costs of hourly matching and abatement (Figure 2B), where greater mitigation is generally associated with higher costs, particularly for regions with lower-quality renewable resources in the East and South. However, some regions offer opportunities for lower-cost hourly matching and mitigation. These include regions with good wind and solar resources and higher coal generation in the baseline such as MISO and Mountain-S. In contrast, regions with relatively clean grids (or emissions policies) are less well-suited for hourly matching due to lower $CO_2$ impacts such as Pacific, California, or Mountain-N. Figure 2B also suggests that abatement costs and premium for hourly matching are generally higher with greater participation. Total abatement is also higher and reflects how the carbon intensity of marginal $CO_2$ does not necessarily change monotonically (Figure 1A), which is reflected in the literature on renewable and clean electricity portfolio standards [22, 23]. Note that the location of 45V clean hydrogen electricity demand implicitly suggests least-cost regional allocation of CFE demand that matches three-pillar demand, given how locations of electrolytic hydrogen production are endogenous (see Methods).

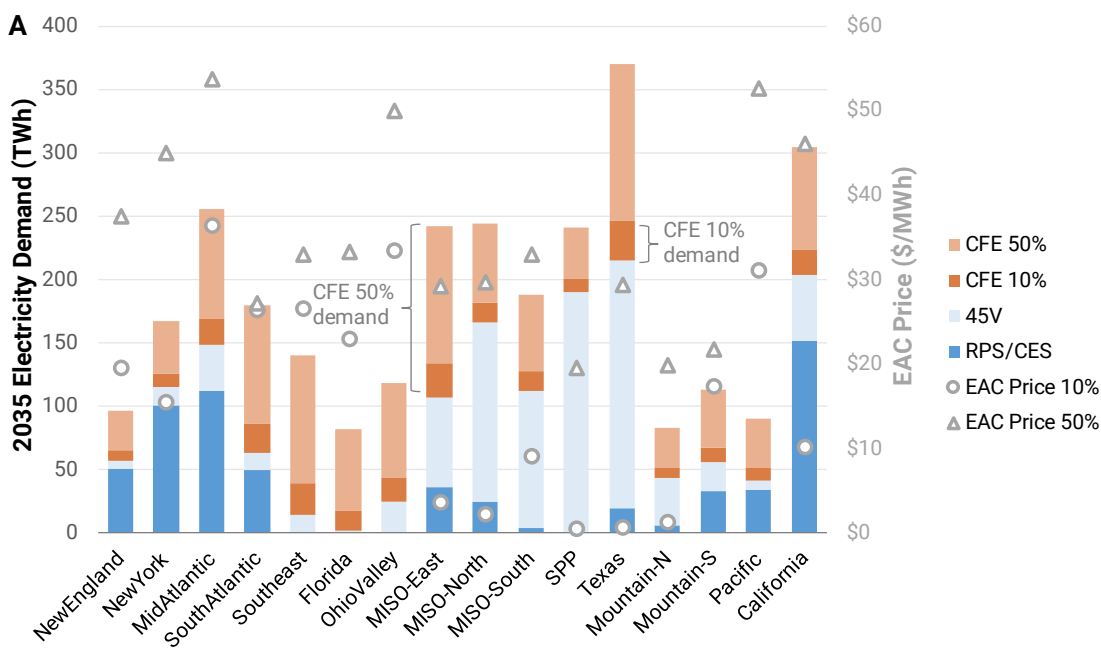

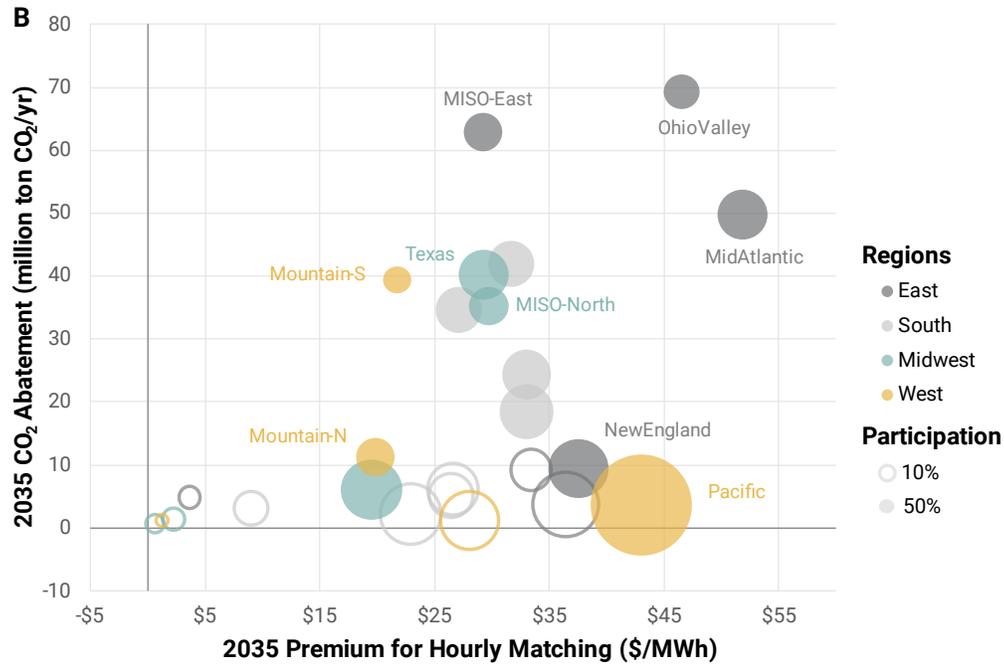

**Figure 2. CFE demand, costs, and trade by region in 2035.** (A) CFE demand by category and EAC price (i.e., shadow prices on the CFE procurement constraint) with three-pillar criteria, where "CFE 50%" shows incremental demand above "CFE 10%" for 50% commercial and industrial procurement. Regional definitions are shown in Figure S1. (B) Comparison of cost premium for hourly matching and abatement for 10% participation scenarios (circles) and 50% participation scenarios (dots). The horizontal axis shows the difference in EAC price with hourly versus annual matching (as opposed to A). Bubble size is proportional to abatement cost. CFE = carbon-free electricity demand; 45V = qualifying generation for electrolysis that receives IRA tax credits for clean hydrogen; RPS/CES = state-level renewable portfolio standards or clean electricity standards; EAC = energy attribute certificate. See Note S3 for abbreviations.

Figure S11 compares magnitudes of EAC prices with wholesale electricity prices across regions and scenarios. For 10% participation, EAC prices are less than half of electricity prices for most regions, except for locations with lower quality wind and solar and consequently higher EAC prices (Figure 2A). For 50% participation, EAC prices increase and are greater than wholesale electricity prices for many regions. However, the greater deployment of low short-run marginal cost resources depresses the wholesale energy prices, and these system changes also may alter capacity prices as well as transmission and distribution costs. These changes have important implications for CFE purchasers as well as other consumers in these markets.

CFE procurement also materially impacts energy storage deployment with cross-regional variation in the magnitude (Figure S12). There is higher energy storage deployment for 50% participation. Increasing the CFE participation rate leads to longer energy storage durations, which increase from 2-7 hours across regions for 10% participation to 5-11 hours for 50% participation. Storage durations are longer for regions with high solar deployment (i.e., in the West and South), while wind-heavy regions have lower energy storage (e.g., SPP, MISO-North).

Differences in capacity mixes and dispatch for illustrative low-cost and high-cost regions are shown in Figure S9 and Figure S10. These results reflect that CFE costs vary significantly across regions, which are driven by differences in regional endowments such as wind and solar resource quality [20, 21]. These regional differences in CFE cost lead to EAC trade with geographical flexibility. Many regions in the East and South are net importers of EACs with relaxed deliverability, while the Midwest and West regions are EAC exporters. These dispatch figures also illustrate the role of inter-regional leakage in compliance with the temporal matching requirement. Since only a portion of load must meet the CFE requirement, EACs can preferentially occupy the bottom of the resource duration curve, and excess CFE generation can displace higher-emitting resources and further reduce emissions on local grids.

*Implications of Available CFE Technologies*

Qualifying CFE resources can have large implications for CFE technology strategy, especially with higher participation rates (regional and national results are shown in Figure 4 and Figure S15, respectively). CFE with three pillars can increase uptake of emerging technological options that may not otherwise be deployed until deeper decarbonization, including advanced nuclear, generation equipped with carbon capture and sequestration (CCS), and longer-duration energy storage. New nuclear is deployed when all zero-emitting technologies qualify as CFE, and gas with CCS is deployed when all low-emitting options can qualify, which is mostly Allam cycle with high $CO_2$ capture rates that requires little $CO_2$ removal to offset its emissions (CCS also benefits from IRA tax credits). Note that many technologies could play this functional role for low-emitting dispatchable technologies, depending on their cost and performance [24]. Nevertheless, the majority of EACs in many regions and scenarios comes from wind, solar, and battery storage.

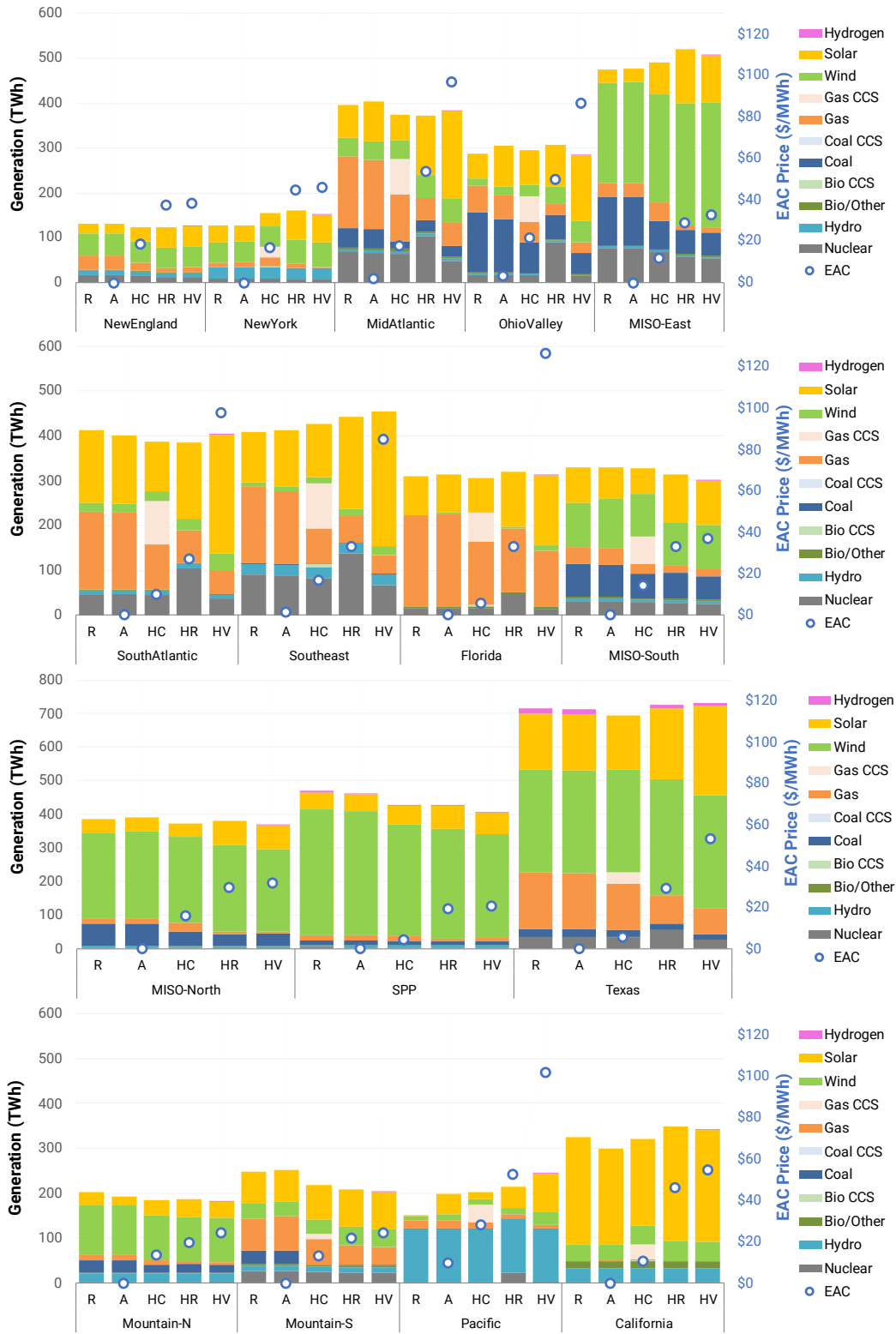

**Figure 3. Regional generation and energy attribute certificate (EAC) prices under alternate CFE demand and technological assumptions (assuming 50% participation rate).** Regional definitions are shown in Figure S1. Scenarios show the reference without CFE demand (R), annual matching (A), hourly

matching with all technologies including CCS (HC), hourly matching with reference technologies (HR), and hourly matching with variable renewables and battery storage only (HV). EAC prices are the shadow prices on the CFE procurement constraint. CCS = carbon capture and sequestration.

Cost increases with limited technological portfolios are highest for regions in the East and South with poor solar and wind endowments (Figure 4). Broader technological portfolios ("All") lead to the lowest EAC prices due to the lower investments needed to reach CFE procurement goals (Figure S13), while the constrained portfolios with VRE and batteries have the highest prices. There is up to a $18/MWh increase between the limited and advanced portfolios with 10% participation and up to $120/MWh with 50% participation. The regions with the highest cost differentials between the limited and advanced technological cases are ones in the Eastern U.S. and Pacific regions that have high EAC prices in the "VRE Only" case in Figure 4, which is also reflected in their lower renewables shares in Figure 3. The smallest differences between technology scenarios are in the Midwest, which are wind-rich regions where the ability to use CCS does not materially alter decisions. Increasing the CFE participation rate also increases costs, especially with restrictions on qualifying technologies.

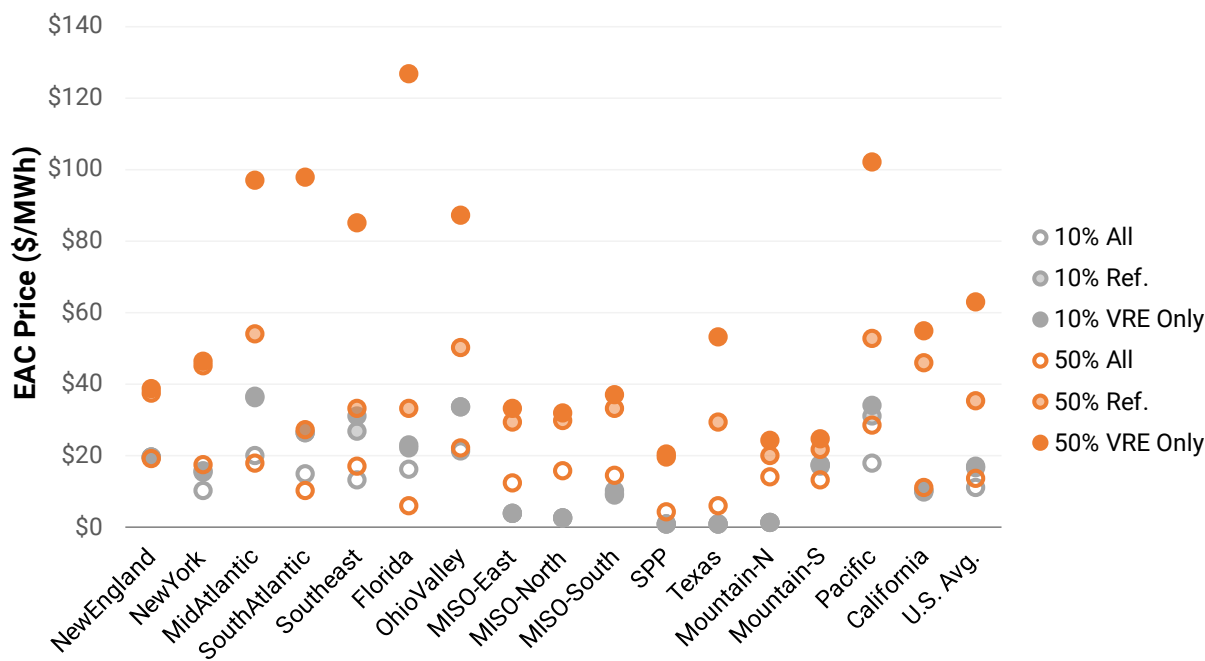

**Figure 4. Energy attribute certificate (EAC) price impacts of CFE demand under alternate technological assumptions with three-pillar criteria.** EAC prices are the shadow prices on the CFE procurement constraint. VRE = variable renewable energy and batteries only.

These EAC prices are comparable to values reported in Xu, et al. [6] for similar U.S. regions. For hourly matching, Xu, et al. find cost differences in California of about $15-25/MWh across technology sensitivities compared to $10-55/MWh here (where the higher end is driven by the higher participation case with limited technologies). Similarly, Xu, et al. find about $7-14/MWh for Wyoming and Colorado, which is similar to $1-24/MWh here for the Mountain-N region. However, our analysis finds broader ranges of regional EAC prices with much higher cost premiums for other regions with lower quality wind and solar resources, especially with more limited technological portfolios.

*Impact of Regional Definitions on Deliverability*

The limited emissions impact when local delivery not enforced (Figure 1A) raises questions about the appropriate size of deliverability regions to reduce emissions without considerably raising costs. For these experiments, scenarios are run with EAC balancing within four large regions (Figure S1) instead of 16 regions, which are similar to the regions specified in the U.S. IRA guidance for clean hydrogen tax credits. These four regions segment the country into the East, South, Midwest, and West, where the West is similar to the Western Interconnection (with the country's best solar resources, per Figure S2) and Midwest states are grouped with Texas (with the country's best wind resources).

Results in Figure S14 illustrate that spatial flexibility in EAC exchange across the four larger regions maintains similar generation changes and emissions reductions as 16 regions. Under 50% participation, national average EAC prices with four regions are the same as 16 regions ($32/MWh), which are higher than the scenario without deliverability ($15/MWh). It not necessarily the size of the region that matters for emissions outcomes but preventing EAC exchange from regions with high CFE development in the reference scenario. As discussed in earlier sections, excess EACs in the baseline come primarily from regions in the West and Midwest that have good renewable resource endowments (making adoption economic in the absence of policy or voluntary procurement) or policies such as binding state-level emissions caps or technology mandates.

*Sensitivities to the Modeling Framework*

Which features are important in a modeling framework to assess CFE procurement strategies? This section examines the impacts of weather years, temporal resolution, and load profiles.

**Weather years:** Earlier results use 2015 meteorology for hourly time-series variables, including potential wind and solar output. This section tests the robustness of results to inter-annual variability by using alternate weather year data from 1999 through 2019, where the capacity mix and dispatch are reoptimized for each weather year. Figure 5 illustrates changes in the installed capacity mix and costs across different weather years. National EAC prices range from $31-37/MWh across weather years, where the 2015 meteorology has among the highest values at $35/MWh, which suggests that this default weather year is challenging for CFE procurement in terms of renewable output (Figure S16 shows above-average drought events for wind regions in 2015).

Although investments and costs for CFE procurement are relatively similar across weather years at a national level, technology-specific shares and regional mixes exhibit greater variability. At a national level, solar, energy storage, and land-based wind have the largest deviations across weather years (Figure 5B). However, these changes can mask larger regional swings in installed capacity (Figure S17). In particular, locations in the East and South with lower-quality renewables vary CFE procurement strategies between solar with storage and new nuclear (Figure S18), though solar and storage are used to some degree regardless of the chosen weather year. Note that these scenarios are conducted for the stringent 50% CFE participation scenario with three pillars, so impacts of weather years would be more limited for other cases.

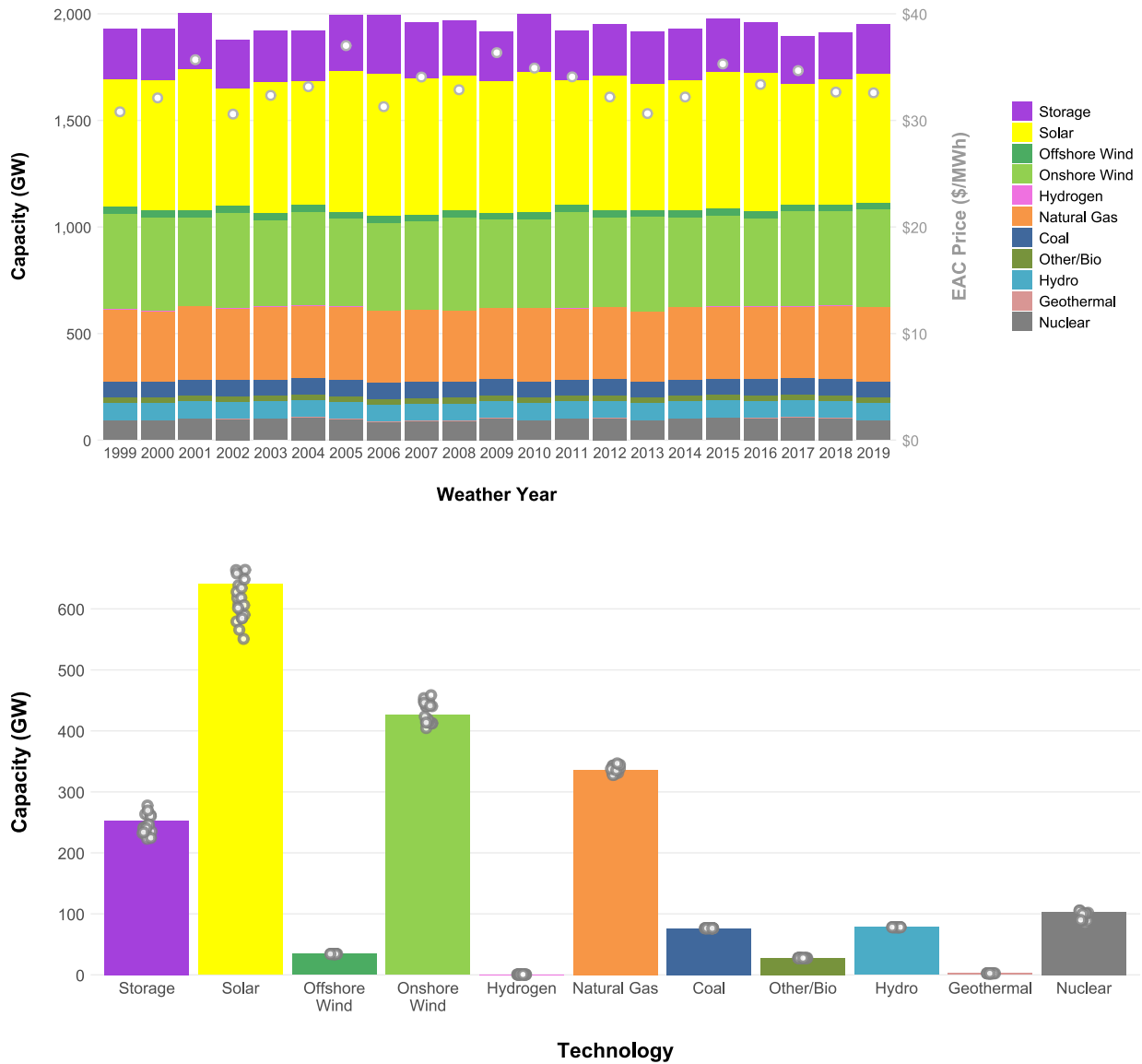

**Figure 5. Impacts of weather year definitions on CFE results for the scenario with three-pillar criteria and 50% participation.** The top panel shows national installed capacity and EAC prices across different weather years. The bottom panel shows technology-specific national installed capacity for the default 2015 weather year (bar) and assuming meteorological conditions from 1999 through 2019 weather years (dots).

**Temporal resolution:** One challenge with representing variable renewables, energy storage, and dispatchable resources is the temporal resolution of a power sector model, which refers to the degree of detail related to time periods within a year. The literature has shown how this choice can materially alter decisions but comes at a large computational cost [25, 26]. Other studies of 24/7 CFE typically use less than full hourly resolution (e.g., Xu, et al. [6] use a "reduced time series of 18 representative weeks" for a single year; Riepin and Brown [8] use a "temporal resolution of 2,920 snapshots"). Earlier results used a version of REGEN with full hourly resolution, and this section uses a reduced-form method of selecting 120 representative segments with chronology [16, 17], which is one example temporal aggregation strategy in the literature [27]. As summarized in Figure S20, these scenarios indicate that full 8,760 hourly resolution leads to greater solar and energy storage deployment for these scenarios and lower wind capacity. These differences are larger for the 50% participation rate than the 10% scenario. EAC prices change by less than 15% across the temporal resolution scenarios.

**Load profiles:** Earlier results use dynamic hourly load profiles that are outputs from REGEN's end-use model [18]. As shown in Figure S7, these profiles exhibit considerable diurnal and seasonal variability, even when aggregated across companies at a regional level. Given how hourly load profiles for CFE procurement are uncertain, this sensitivity uses flat demand profiles (that match the aggregate annual demand from earlier sections) to understand how modeled impacts could change. As shown in Figure S19, assuming flat CFE demand rather than dynamic hourly loads has relatively small impacts on the generation mix with three-pillar criteria. Emissions impacts track generation changes, where load profiles have smaller impacts than voluntary program design decisions (e.g., annual versus hourly matching).

*Sensitivity to the Policy Environment*

Earlier scenarios assumed a background of current federal and state policies and incentives. However, there is uncertainty about whether IRA incentives will remain in place or augmented in future years. This section conducts alternate experiments that remove IRA incentives and add a carbon fee that starts at $20/t-$CO_2$ in 2025 and rises at 3% annually above inflation. These scenarios could alternatively be viewed through the lens of non-U.S. geographical contexts that may not have technological subsidies or that may have carbon pricing.

As shown in Figure S21, both annual and hourly matching are less effective in policy environments with more stringent emissions pricing or deployment incentives. Annual matching has larger impacts vis-à-vis hourly matching in environments that bring less clean energy in the baseline before CFE demand. In other words, a scenario without subsidies has the largest response with annual matching. Similarly, CFE demand with three pillars has the largest impacts in markets without climate policy: 109 million tons of $CO_2$/yr reduction under "No IRA" scenario, 42 under IRA, and 16 under a carbon fee. Annual matching has largest $CO_2$ reductions under the limited climate policy condition, where reductions of 51 Mt-$CO_2$/yr are considerably larger than the <1 Mt-$CO_2$/yr with IRA or carbon fee. These results indicate that annual matching may have been better suited to reduce emissions in earlier policy environments compared to current policies, which have federal tax credits and state decarbonization policies.

**Discussion**

*Conclusions*

This research identifies key challenges in implementing hourly CFE procurement, including technological and market barriers, and offers insights into potential environmental and economic impacts. A key finding from across the scenarios is that the three pillars of hourly matching, incrementality, and deliverability support maximizing emissions reductions from CFE procurement, especially after IRA's passage, which brings more wind and solar in the baseline. Conversely, emissions pledges that do not contain these elements may not achieve claimed reductions, especially if procurement occurs in a country or region with clean electricity subsidies, emissions policies, or technology mandates. The analysis indicates that emissions accounting frameworks based on annual matching may not accurately reflect actual $CO_2$ impacts from procurement, especially for future power systems where policy, technology, and market trends encourage greater shares of clean electricity deployment.

Second, meeting three-pillar qualification criteria increases costs of CFE procurement. These costs vary by region, participation level, and technology availability, spanning $11-63/MWh nationally across scenarios and $1-130/MWh across regions. Note that these scenarios look at cases with 100% temporal matching, and earlier analysis indicates that costs can be reduced for lower matching rates [6, 8, 15]. Large regional differences in emissions and generation responses in our analysis highlight the importance of region-specific assessments and how results in previous studies for regions in Western U.S. are likely not generalizable, given their high-quality renewable resources and state policies. Although this finding suggests that caution is warranted in extrapolating results to other geographies, the breadth of regional conditions studied in the analysis suggests that broad insights may be transferrable to other countries or subnational jurisdictions if conditions are sufficiently similar (e.g., emissions and technology policies, resource endowments, fuel prices, existing capacity and infrastructure, technological costs). These scenarios also highlight how regional trade dynamics and spillover effects can alter emissions and costs of CFE procurement, which underscores the importance of representing neighboring systems.

Third, this analysis underscores the importance of advanced technologies for managing costs, especially having broader technological portfolios for regions with lower renewable resource quality. Allowing broader technology portfolio to qualify for CFE procurement (including CCS) can lower costs of three-pillar procurement up to 57% under 10% participation and 96% under 50% participation. As others have noted [15], 24/7 CFE procurement could accelerate electricity decarbonization through induced technological learning and helping emerging energy technologies to become more cost-competitive, which can create a virtuous cycle of advancing innovation, accelerating deployment, and lowering project risk. The analysis indicates that three-pillar CFE can provide early market opportunities for advanced technologies, including low-emitting dispatchable/firm generation and long-duration energy storage, which reinforces earlier analysis [15, 8, 14]. Since technological learning effects may have diminishing marginal returns, early projects could have relatively large impacts on commercialization [15]. The results also suggest that energy storage is a cornerstone, especially for hourly matching, and gains importance for deeper decarbonization.

Fourth, the analysis indicates that effects of CFE procurement depend on interactions with other existing policies and incentives for low-emitting electricity. More stringent policies lead to less "additional" clean electricity and more limited emissions reductions than in the absence of policies. Annual matching may be more suitable for reducing emissions in geographies and times without subsidies or with higher relative costs for low-emitting electricity, both of which are less common in many current conditions. More broadly, the analysis underscores how CFE impacts are contingent on assumptions about the future, including changes in policy, technology, and markets, which are fundamentally uncertain.

Finally, this analysis highlights how the modeling framework can influence insights about the costs and technology impacts of CFE procurement. We demonstrate the impacts that assumed weather years, temporal resolution for intra-annual segments, and load shapes can affect CFE procurement. These effects are dependent on the scenario, region, and output of interest.

*Future Work*

The analysis identifies several areas for future work. First, these scenarios exhibit different levels of decarbonization but do not reach the goal of economy-wide net-zero emissions. Future work can quantify how impacts of CFE procurement could differ when targeting deep decarbonization under different policy drivers [28, 29]. Second, the analysis examines aggregate CFE demand and not trading across entities with different load shapes, which essentially assumes a liquid market for time-based EAC trading. Future work can look at multilateral trading in EAC markets with entities with distinct load profiles, building on earlier analysis [30]. Third, the analysis assumes exogenous CFE targets for each model region. However, some CFE loads may have endogenous locational decisions and load flexibility, which may affect costs and emissions impacts, potentially including data centers [31]. Fourth, in addition to the uncertainties discussed earlier, omitted dynamics imply that the results should not be viewed as predictions but rather as scenarios that provide insight across a range of conditions. Two important omitted dynamics that would be good areas for future study are intra-regional grid congestion and interconnection queues [32]. Fifth, this analysis illustrates the sensitivity of the deterministic optimization of CFE procurement to the choice of a single weather year; however, these scenarios do not necessarily inform how optimal decisions can be made robust to weather uncertainty, which is an important area for future research. Finally, the results show how moving from annual matching to hourly CFE leads to significantly more energy storage, which raises questions about emissions accounting with energy storage. These are a subset of larger trends across the power sector and energy systems that may influence CFE procurement, including increased end-use electrification, deployment of distributed energy resources, role of emerging supply- and demand-side options, and growing loads from manufacturing and data centers.

**Methods**

*Model*

To examine the effects of clean energy procurement on regional power systems, this analysis uses EPRI's U.S. Regional Economy, Greenhouse Gas, and Energy (REGEN) model, which is a state-of-the-art model of energy systems that has been applied across a range of peer-reviewed studies, model intercomparisons, and technical reports.

REGEN's electric sector model is an intertemporal capacity planning and dispatch model that makes simultaneous decisions about investments and retirements, transmission, and system operations with hourly correlations between load, wind output, and solar output [19]. The optimization model determines the least-cost mix of resources given assumptions about technology costs, markets, and policies. This version of the model uses full hourly temporal resolution with investments and operations for a single future period (2035), which helps to better characterize the economics of energy storage and balancing resources while remaining computational tractable. Sensitivities include an intertemporal optimization that uses five-year time periods through 2050 with 120 representative hours per year with a novel method of representing chronology between these individual periods [16, 17]. REGEN represents 16 interconnected regions in the continental U.S. with transmission expansion and hourly trade (Figure S1). The electric sector and fuels model is formulated as a large-scale linear optimization with a single decision-maker with perfect foresight [33, 34] that minimizes the net present value of system costs subject to technical, economic, and policy constraints.

Hourly regional electricity profiles are outputs from the REGEN end-use model, which has sector-specific technological deployment and hourly electricity use [18]. Hourly load time-series of load varies regionally based on climate, existing building and technology stocks, projected end-use changes, industrial composition, and other factors. Figure S6 illustrates hourly electricity demand for two regions and time periods across different end uses, and Figure S7 shows an illustrative aggregate CFE load profile for New England in 2035. Note that this analysis aggregates CFE demand at a regional level across commercial and industrial companies, which implicitly assumes a liquid market for time-based EAC trading. Other work in the literature has looked at multilateral trading in EAC markets with entities with distinct load profiles [30].

The hourly voluntary CFE market-clearing constraint under conditions where EACs must coincide temporally and spatially with production from qualified resources is:

$$\sum_{i \in I} X_{ihrt} + \sum_{j \in J} [D_{jhrt} - \rho_j C_{jhrt}] \geq d_{hrt} \qquad \forall h, r, t \qquad (1)$$

Where $X_{ihrt}$ is generation from eligible technologies $i$ in hour $h$, region $r$, and time period $t$. $D_{jhrt}$ is the discharge from energy storage technology $j$, and $C$ is charge with penalty $\rho_j$. The analysis assumes that excess generation of qualified CFE resources can be curtailed, stored, or sold to the regional market at wholesale prices. Shadow prices on the CFE constraint give EAC prices, which represent the cost premium of CFE procurement relative to least-cost electricity procurement. Since EAC prices can vary across hours and regions depending on the scenario (e.g., 24/7 CFE creates hourly differentiated EAC products for each region), results use the generation-weighted average of EAC prices to aggregate across regions and over the year.

Detailed documentation of the model and datasets can be found at: https://us-regen-docs.epri.com/

*Scenario Design*

The reference scenario includes all on-the-books federal and state electric sector policies and incentives, including the Inflation Reduction Act (IRA) but no explicit national $CO_2$ policy for the power sector or economy. Policies included in the reference and all other scenarios include state-level renewable portfolio standards, clean electricity standards, technology-specific mandates (e.g., offshore wind, energy storage, solar carve-outs), carbon pricing (e.g., California's economy-wide cap-and-trade, Regional Greenhouse Gas Initiative $CO_2$ caps), and nuclear moratoria. The analysis does not include U.S. Environmental Protection Agency regulations on power plants, given their political uncertainty [35].

IRA incentives are included for the electric sector, end-use sectors, and low-emitting energy supply [36, 37]. Key IRA provisions include:

- **45Y Clean Electricity Production Credit (IRA §13701):** Projects receive up to $30/MWh for 10 years, including an endogenous representation of energy community bonuses (Figure S3), which is technology neutral beginning in 2025 for all technologies with "emissions intensity not greater than zero."
- **48E Clean Electricity Investment Credit (IRA §13702):** Projects receive a 30% credit with labor bonus and 10 percentage point bonuses for domestic content and energy communities (the domestic content bonus is not included here). REGEN allows technologies in different regions to endogenously select between the production and investment tax credits.
- **45Q $CO_2$ Capture and Storage Credit (IRA §13104):** Projects receive up to $85/t-$CO_2$ captured with the labor bonus. There is a 12-year eligibility for projects, which must commence construction by 2032. Like 45V, there are not domestic content or energy communities bonuses.
- **45V Clean Hydrogen Production Credit (IRA §13204):** The clean hydrogen subsidy schedule depends on the lifecycle emissions intensity of production, up to $3/kg with 10-year eligibility (must begin construction by 2032). IRA credits for clean hydrogen include Treasury guidance with "three pillars" criteria. These scenarios use endogenous uptake of 45V credits and endogenous location and operational decisions for electrolytic hydrogen production, which means that the regional allocation in Figure 2A represents the cost-minimizing mix.

REGEN represents a range of existing and emerging electricity generation technologies, and capital cost assumptions over time for a subset of these options are shown in Figure S4. REGEN also includes a variety of energy storage technologies such as short- and long-duration batteries (with endogenous durations), compressed air energy storage, electrolytic hydrogen, and existing pumped hydro.

Scenarios assume CFE demand is a share of commercial and industrial load (Figure S5). Changing electricity demand and load shapes are outputs from REGEN's end-use model (Figure S6).

In the technological sensitivities that allow all technologies, including CCS balanced by carbon removals, the modeling adds a constraint on $CO_2$ emissions for the voluntary CFE market. The constraint requires that national emissions from qualified resources must reach net-zero $CO_2$ levels on an annual basis. Carbon removal to balance residual emissions can come from bioenergy with CCS in the power sector or direct air capture.

*Caveats*

There are several caveats to bear in mind when interpreting the results:

- This analysis examines aggregate CFE demand profiles for each model region (Figure S7) and not trading across entities with different load shapes.
- The scenarios primarily hold all other climate and energy policies constant across scenarios and do not look at long-run federal $CO_2$ policy (e.g., to reach net-zero emissions by 2050), though Figure S21 illustrates impacts of a power sector carbon fee on CFE procurement.
- The analysis uses the same hourly load shapes across all scenarios for comparability.

**Acknowledgments**

The views and opinions expressed in this paper are those of the authors alone and do not necessarily reflect those of EPRI or its members.

**Code and Data Availability**

All data associated with the analysis are available at https://epri.box.com/s/d7z0x22loavrhjmbysup9mz54p4ng59f. The analysis uses the U.S. Regional Economy, Greenhouse Gas, and Energy (REGEN) model, and detailed documentation, other peer-reviewed articles and reports, as well as source code for the electric sector and fuels model are available at https://esca.epri.com/usregen/.

**Competing Interests**

The authors declare no competing interests.

**References**

[1] EPRI, "24/7 Carbon-free Energy: Matching Carbon-free Energy Procurement to Hourly Electric Load," EPRI, Palo Alto, CA, 2023.

[2] Google, "24/7 by 2030: Realizing a Carbon-Free Future," 2020.

[3] 24/7 Carbon-Free Energy Compact, "24/7 Carbon-Free Energy: Methods, Impact, and Benefits," 2021.

[4] The White House, "Executive Order on Catalyzing Clean Energy Industries and Jobs Through Federal Sustainability," Washington, DC, 2021.

[5] Greenhouse Gas Protocol, "Standards Update Process: Frequently Asked Questions," 2023.

[6] Q. Xu, W. Ricks, A. Manocha, N. Patankar and J. Jenkins, "System-Level Impacts of Voluntary Carbon-Free Electricity Procurement Strategies," *Joule,* vol. 8, no. 2, pp. 374-400, 2024.

[7] AES, "24/7 Carbon-Free Energy," 2021.

[8] I. Riepin and T. Brown, "On the Means, Costs, and System-Level Impacts of 24/7 Carbon-Free Energy," *Energy Strategy Reviews,* vol. 54, p. 101488, 2024.

[9] G. Blanford and J. Bistline, "Impacts of IRA's 45V Clean Hydrogen Production Tax Credit," EPRI, Palo Alto, CA, 2023.

[10] W. Ricks, Q. Xu and J. Jenkins, "Minimizing Emissions from Grid-Based Hydrogen Production in the United States,"

*Environmental Research Letters,* vol. 18, no. 1, p. 014025, 2023.

[11] M. Giovanniello, A. Cybulsky, T. Schittekatte and D. Mallapragada, "The Influence of Additionality and Time-Matching Requirements on the Emissions from Grid-Connected Hydrogen Production," *Nature Energy,* vol. 9, no. 2, pp. 197-207, 2024.

[12] D. Schlund and P. Theile, "Simultaneity of Green Energy and Hydrogen Production: Analysing the Dispatch of a Grid-Connected Electrolyser," *Energy Policy,* vol. 166, p. 113008, 2022.

[13] O. Ruhnau and J. Schiele, "Flexible Green Hydrogen: The Effect of Relaxing Simultaneity Requirements on Project Design, Economics, and Power Sector Emissions," *Energy Policy ,* vol. 182, p. 113763, 2023.

[14] IEA, "Advancing Decarbonisation Through Clean Electricity Procurement," International Energy Agency, Paris, France, 2022.

[15] I. Riepin, J. Jenkins, D. Swezey and T. Brown, "24/7 Carbon-Free Electricity Matching Accelerates Adoption of Advanced Clean Energy Technologies," *Joule,* 2025.

[16] G. Blanford, J. Merrick, J. Bistline and D. Young, "Simulating Annual Variation in Load, Wind, and Solar by Representative Hour Selection," *The Energy Journal,* vol. 39, no. 3, p. 183–207, 2018.

[17] J. Merrick, J. Bistline and G. Blanford, "On Representation of Energy Storage in Electricity Planning Models," *Energy Economics,* vol. 136, p. 107675, 2024.

[18] J. Bistline, C. Roney, D. McCollum and G. Blanford, "Deep Decarbonization Impacts on Electric Load Shapes and Peak Demand," *Environmental Research Letters,* vol. 16, no. 9, 2021.

[19] EPRI, "US-REGEN Documentation," EPRI, Palo Alto, CA, 2023.

[20] J. Bistline, M. Browning, J. DeAngelo, D. Huppmann, R. Jones, J. McFarland, A. Molar-Cruz, S. Rose and S. J. Davis, "Uses and Limits of National Decarbonization Scenarios to Inform Net-Zero Transitions," *Joule,* 2024.

[21] S. Rose and A. Molar-Cruz, "Differences in Regional Decarbonization Opportunities, Uncertainties, and Risks," EPRI, Palo Alto, CA, 2023.

[22] D. Young and J. Bistline, "The Costs and Value of Renewable Portfolio Standards in Meeting Decarbonization Goals," *Energy Economics,* vol. 73, pp. 337-351, 2018.

[23] L. Goulder, M. Hafstead and R. Williams III, "General Equilibrium Impacts of a Federal Clean Energy Standard," *American Economic Journal: Economic Policy,* vol. 8, no. 2, pp. 186-218, 2016.

[24] N. Sepulveda, J. Jenkins, F. de Sisternes and R. Lester, "The Role of Firm Low-Carbon Electricity Resources in Deep Decarbonization of Power Generation," *Joule,* vol. 2, no. 11, pp. 2403-2420, 2018.

[25] J. Bistline, "The Importance of Temporal Resolution in Modeling Deep Decarbonization of the Electric Power Sector," *Environmental Research Letters,* vol. 16, no. 8, p. 084005, 2021.

[26] H. Teichgraeber and A. Brandt, "Time-Series Aggregation for the Optimization of Energy Systems: Goals, Challenges, Approaches, and Opportunities," *Renewable and Sustainable Energy Reviews,* vol. 157, p. 111984, 2022.

[27] M. Hoffmann, J. Priesmann, L. Nolting, A. Praktiknjo, L. Kotzur and D. Stolten, "Typical Periods or Typical Time Steps? A Multi-Model Analysis to Determine the Optimal Temporal Aggregation for Energy System Models," *Applied Energy,* vol.

304, p. 117825, 2021.

[28] M. Browning, J. McFarland, J. Bistline, G. Boyd, M. Muratori, M. Binsted, C. Harris, T. Mai, G. Blanford, J. Edmonds, A. Fawcett, O. Kaplan and J. Weyant, "Net-Zero CO2 by 2050 Scenarios for the United States in the Energy Modeling Forum 37 Study," *Energy and Climate Change,* vol. 4, p. 100104, 2023.

[29] J. DeAngelo, I. Azevedo, J. Bistline, L. Clarke, G. Luderer, E. Byers and S. Davis, "Energy Systems in Scenarios at Net-Zero CO2 Emissions," *Nature Communications,* 2021.

[30] Q. Xu and J. Jenkins, "Electricity System and Market Impacts of Time-based Attribute Trading and 24/7 Carbon-free Electricity Procurement," Zenodo, 2022.

[31] EPRI, "Powering Data Centers: U.S. Energy System and Emissions Impacts of Growing Loads," EPRI, 2024.

[32] L. Armstrong, A. Canaan, C. Knittel, G. Metcalf and T. Schittekatte, "Can Federal Grid Reforms Solve the Interconnection Problem?," *Science,* vol. 385, no. 6704, pp. 31-33, 2024.

[33] M. Hoffmann, B. Schyska, J. Bartels, T. Pelser, J. Behrens, M. Wetzel, H. Gils, C. Tang, M. Tillmanns, J. Stock and A. Xhonneux, "A Review of Mixed-Integer Linear Formulations for Framework-Based Energy System Models," *Advances in Applied Energy,* p. 100190, 2024.

[34] J. Merrick, "Analysis of Foresight in Long-Term Energy System Models," EPRI, Palo Alto, CA, 2021.

[35] J. Bistline, A. Bergman, G. Blanford, M. Brown, D. Burtraw, M. Domeshek, A. Fawcett, A. Hamilton, G. Iyer, J. Jenkins, B. King, H. Kolus, A. Levin, Q. Luo, K. Rennert, M. Robertson, N. Roy, E. Russell, D. Shawhan, D. Steinberg and A. v. Brummen, "Impacts of EPA's Finalized Power Plant Greenhouse Gas Standards," *Science,* vol. 387, no. 6730, pp. 140-143, 2025.

[36] J. Bistline, G. Blanford, M. Brown, D. Burtraw, M. Domeshek, J. Farbes, A. Fawcett, A. Hamilton, J. Jenkins, R. Jones, B. King, H. Kolus, J. Larsen, A. Levin, M. Mahajan, C. Marcy, E. Mayfield, J. McFarland, H. McJeon, R. Orvis and Patank, "Emissions and Energy Impacts of the Inflation Reduction Act," *Science,* vol. 380, no. 6652, pp. 1324-1327, 2023.

[37] J. Bistline, M. Brown, M. Domeshek, C. Marcy, N. Roy, G. Blanford, D. Burtraw, J. Farbes, A. Fawcett, A. Hamilton, J. Jenkins, R. Jones, B. King, H. Kolus, J. Larsen, A. Levin, M. Mahajan, E. Mayfield, J. McFarland, H. McJeon, R. Orvis and N. Patankar, "Power Sector Impacts of the Inflation Reduction Act of 2022," *Environmental Research Letters,* vol. 19, no. 1, p. 014013, 2024.

[38] EPRI, "Program on Technology Innovation: Generation Technology Options - 2024," EPRI, Palo Alto, CA, 2024.

**Supplementary Information**

**System Effects of Carbon-Free Electricity Procurement: Regional Technology and Emissions Impacts of Voluntary Markets**

John Bistline[1]*, Geoffrey Blanford[1], Adam Diamant[1], Arin Kaye[1], Daniel Livengood[1], Qianru Zhu[1]

**Affiliations:**

[1] Electric Power Research Institute; Palo Alto, USA.
*Corresponding author. Email: jbistline@epri.com.

**Supplementary Note 1: Overview of Methods and Scenario Assumptions**

Figure S1 shows the regional definitions for this analysis.

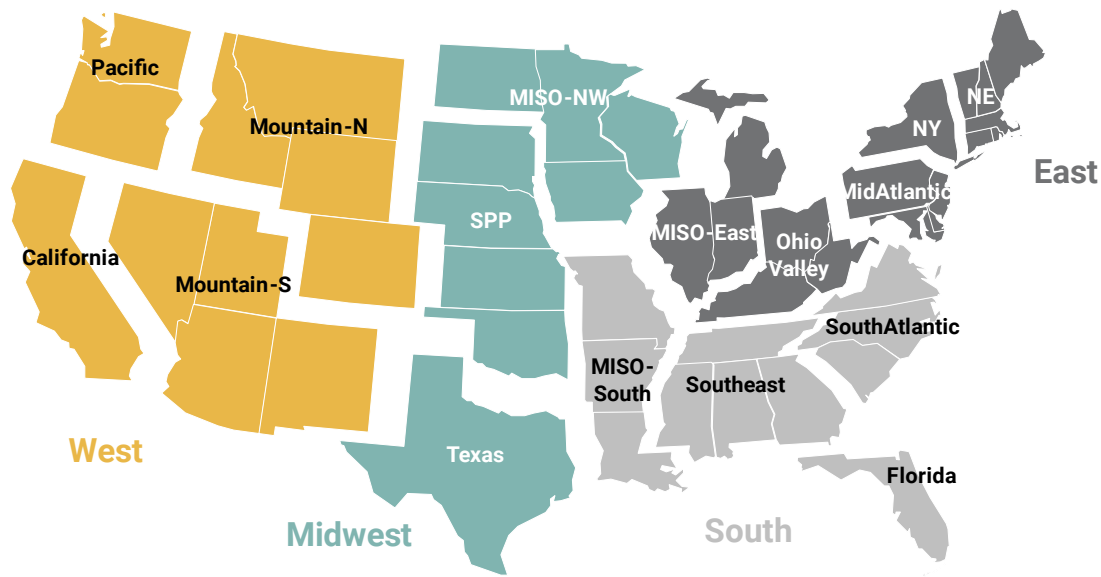

**Figure S1. Map of regional configuration for this study.** Capacity planning and dispatch decisions occur for each of these 16 regions. REGEN four reporting regions are also shown.

Figure S2 illustrates wind and solar resource maps. Detailed discussions of wind and solar resources and their hourly profiles in REGEN are provided in the model documentation.

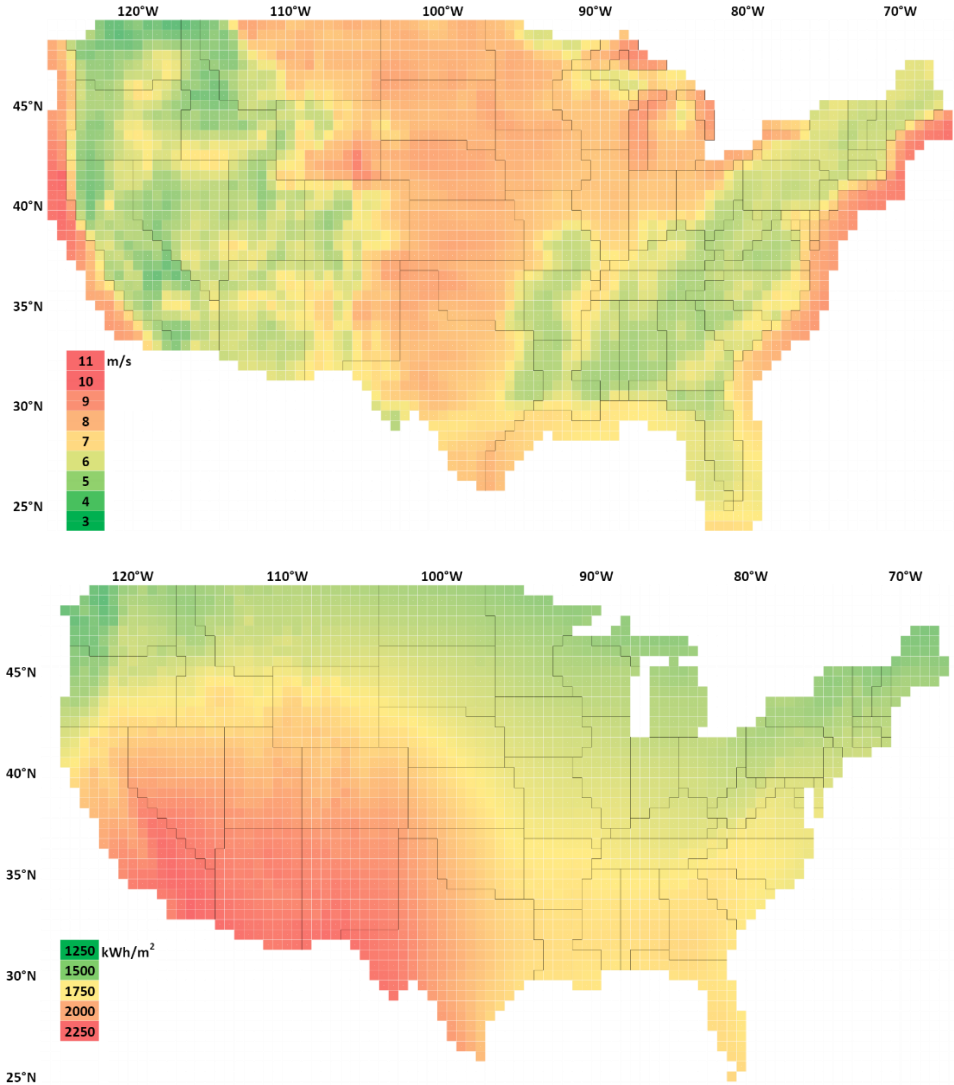

**Figure S2. Resource maps for wind (top panel) and solar (bottom panel).** The wind map shows the long-run average (from 1980 through 2015) wind speed at 100 meters by grid cell based on NASA's MERRA-2 reanalysis dataset. The solar map shows the long-run average annual Global Horizontal Irradiance (GHI) based on NASA's MERRA-2 reanalysis dataset.

For IRA's production and investment tax credits, the modeling provides endogenous selection between credit types and the energy communities bonus, which provides an additional 10% for the production credit and 10 percentage points for the investment credit. Figure S3 shows the areas that qualify based on criteria related to coal mine or power plant closures as well as fossil fuel employment.

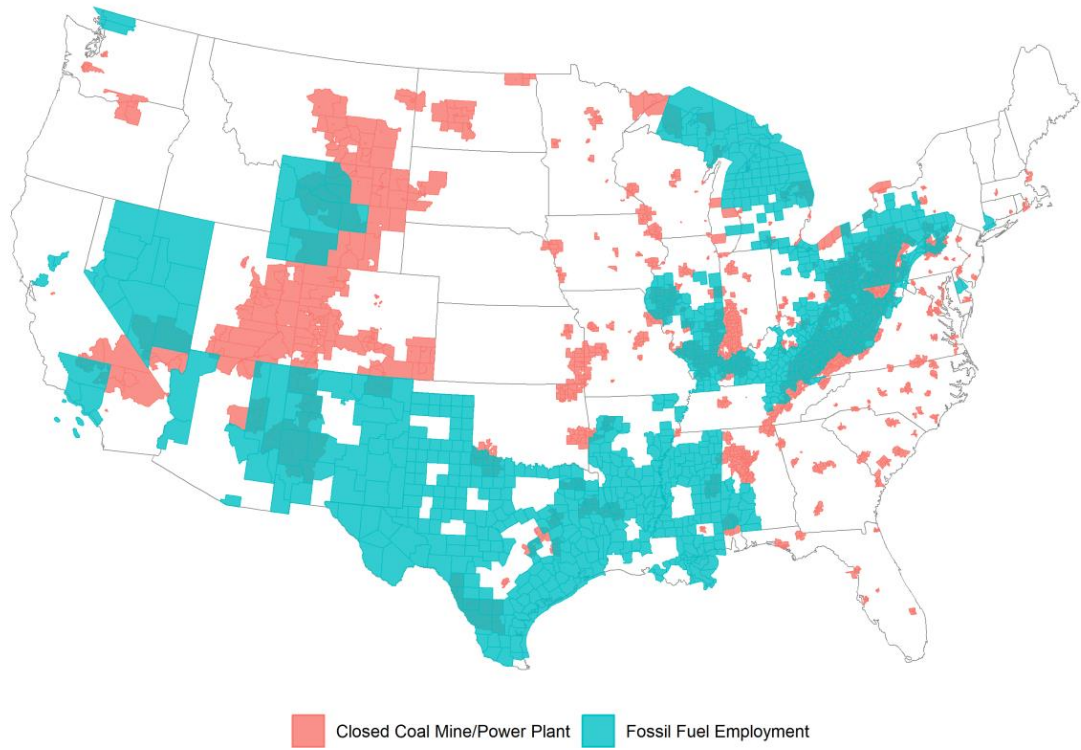

**Figure S3. Map of the areas qualifying for the energy communities bonus under the Inflation Reduction Act.** Based on U.S. Department of Energy's energy communities definition (link).

Assumed capital costs over time for select generation technologies are shown in Figure S4. Technological cost and performance assumptions are based on EPRI's Technology Assessment Guide [38] and are summarized in the REGEN documentation site: https://us-regen-docs.epri.com/.

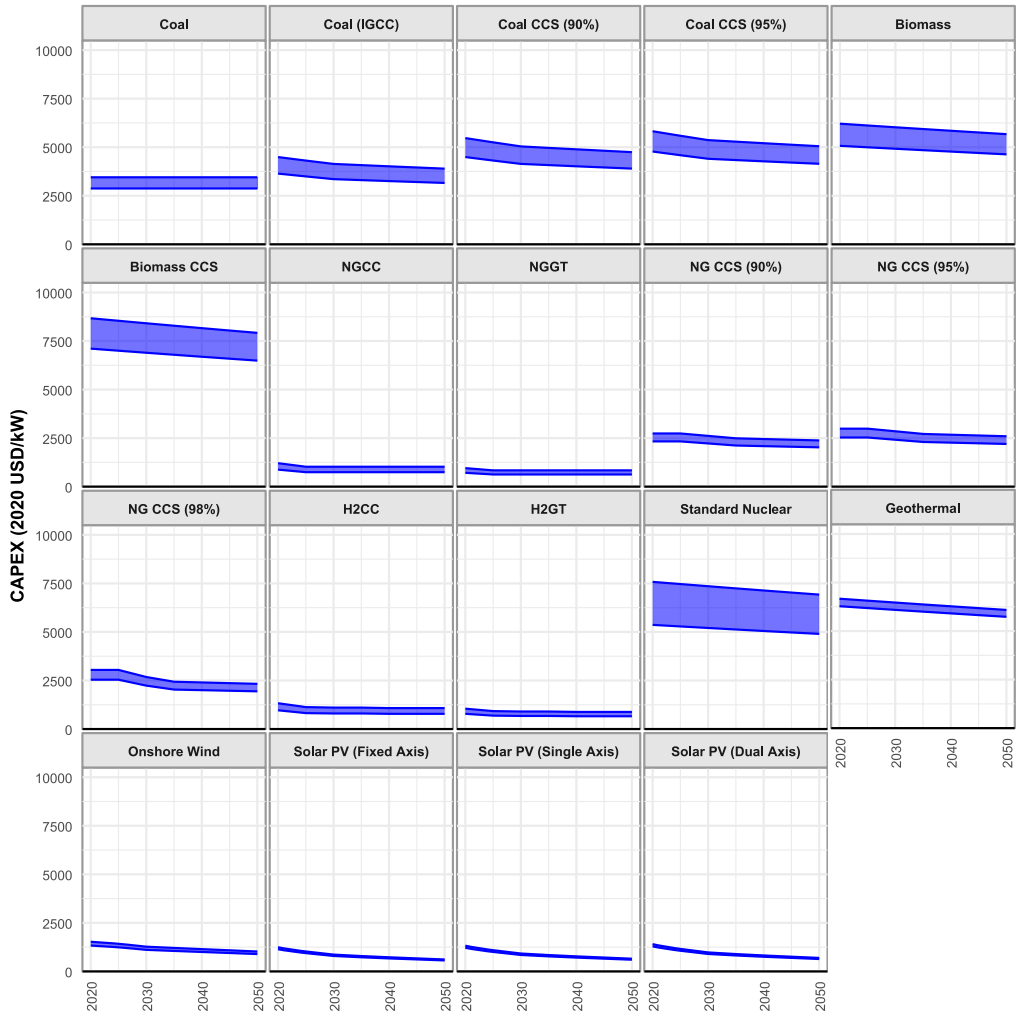

**Figure S4. Capital cost assumptions for electricity generation technologies over time.** Ranges show variation across model regions.

Commercial and industrial electricity demand over time comes from REGEN's end-use model (Figure S5). CFE demand is a share of commercial, industrial, and transport load (since commercial vans, trucks, and other electrified vehicles owned by companies with 24/7 CFE pledges). Note that the focus year for this analysis—2035—has extensive electrolytic hydrogen demand from 45V incentives under IRA, which declines after these tax credits expire.

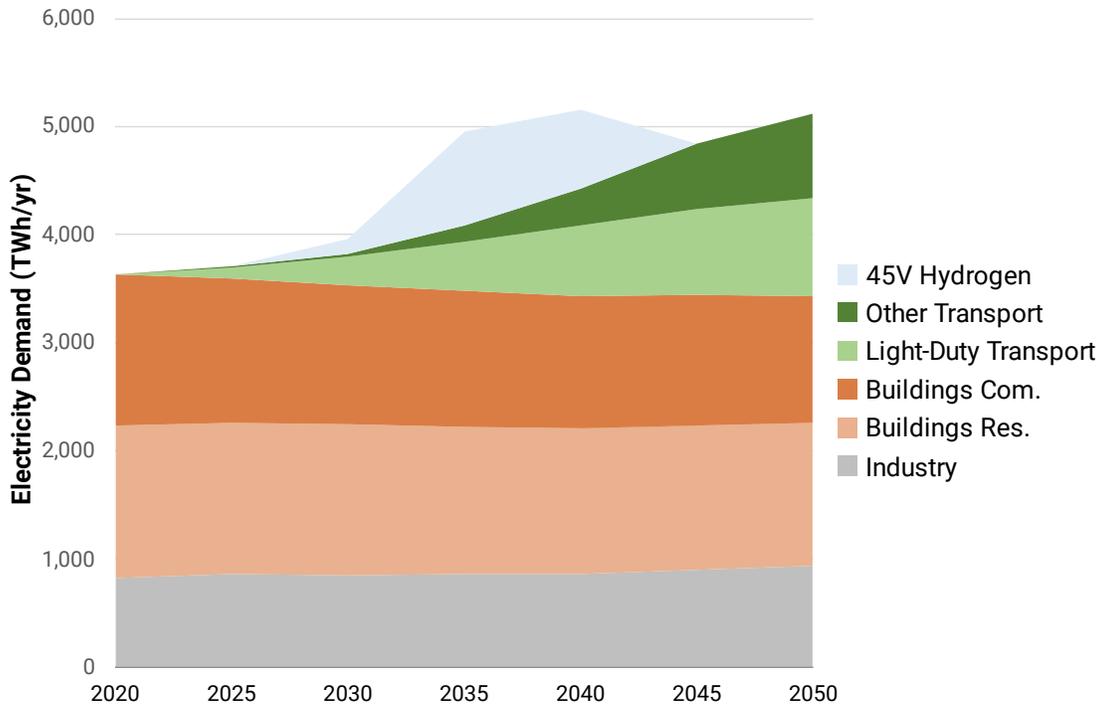

**Figure S5. Electricity demand by sector over time.** Results are outputs from EPRI's REGEN model for a current policies scenario.

As shown in Figure S7, hourly load time-series varies regionally based on climate, existing building and technology stocks, projected end-use changes, industrial composition, and other factors [18]. For instance, New England's winter peak from space heating grows with heat pump deployment by 2035, while the cooling shape is larger in California. Figure S7 shows an example of the CFE demand profile for New England in 2035 with the 10% C&I participation case. These aggregate hourly profiles are composites of REGEN regional hourly end-use profiles and exhibit seasonal and diurnal shapes.

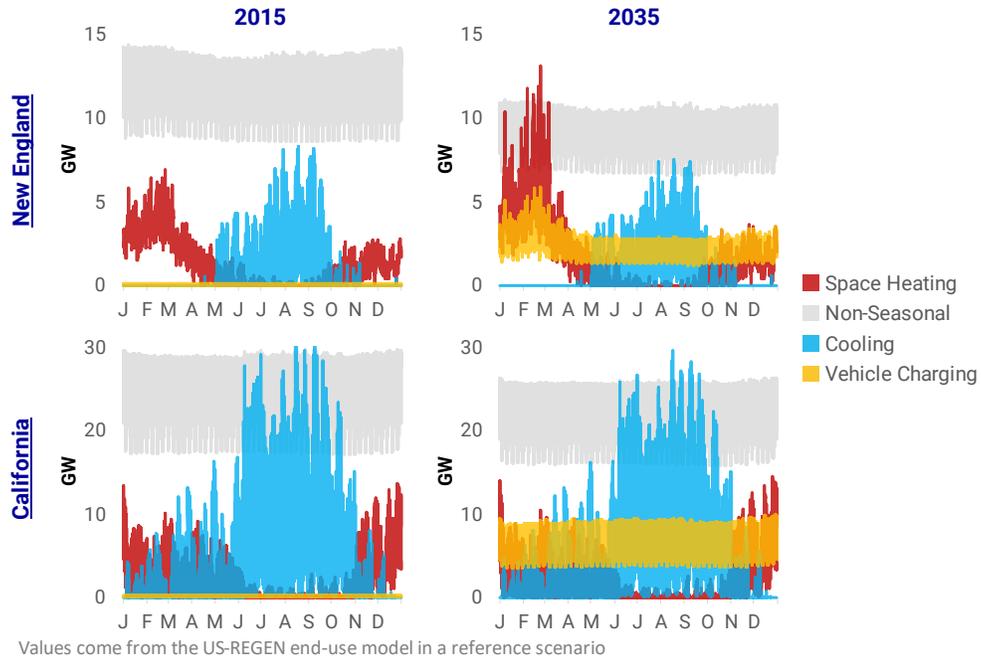

**Figure S6. Hourly electricity demand profiles by end use and region in 2015 (left column) and 2035 (right column).** Results are outputs from EPRI's REGEN model for a current policies scenario.

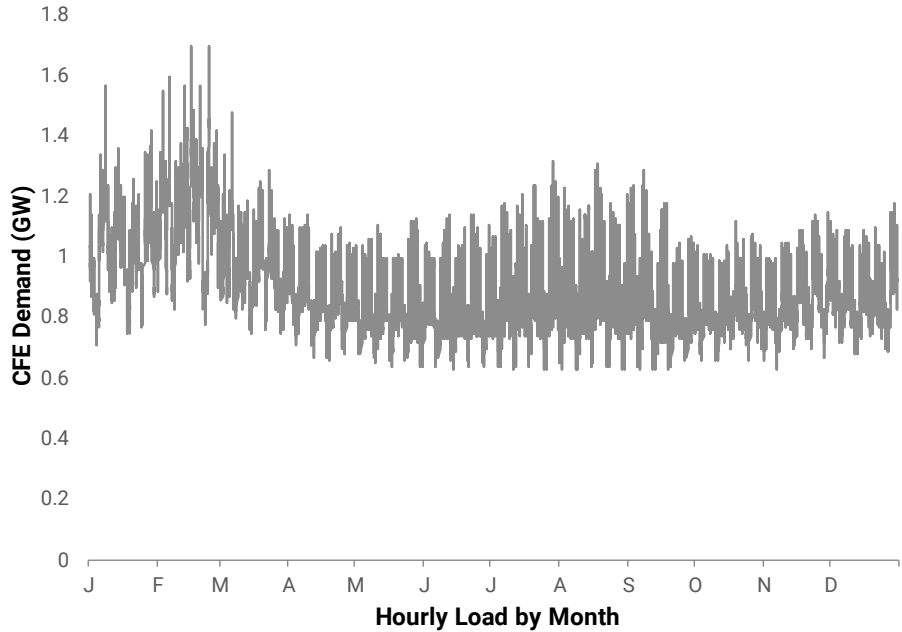

**Figure S7. Hourly CFE demand profile for New England in 2035 (assuming 10% participation).** Results are outputs from EPRI's REGEN model for a current policies scenario.

**Supplementary Note 2: Additional Results**

*Regional Results*

Higher CFE participation leads to lower $CO_2$ "leakage" from relaxed qualification criteria (Figure S8). It also alters the ranking of qualification pillar impacts—enforcing temporal matching is more important with higher CFE demand. Absolute $CO_2$ reductions also increase in CFE demand with three-pillar reductions from 42 Mt-$CO_2$/yr to 466 moving from 10% rate to 50% rate. This reflects more coal displaced at the margin relative to natural gas (Figure 1).

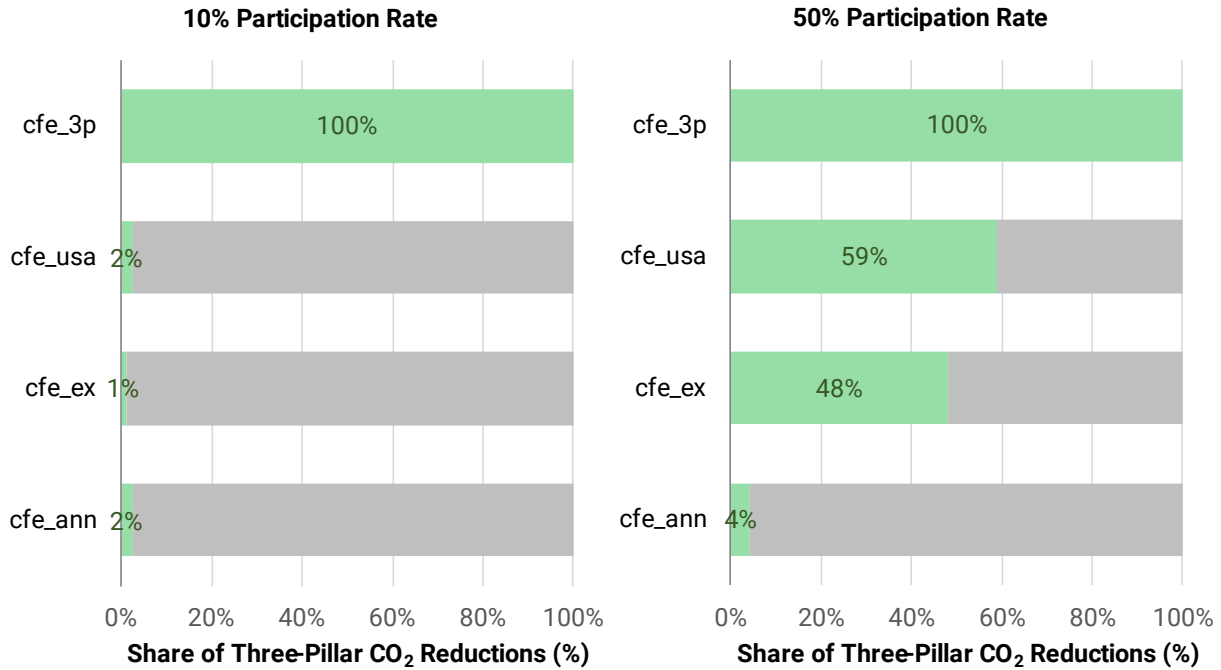

**Figure S8. $CO_2$ reductions by scenario in 2035 relative to the three-pillar case.** Changes with 10% C&I and 50% CFE demand are shown (left and right panels, respectively).

Regional dispatch dynamics are shown below for a week in SPP (Figure S9) under the three-pillar scenarios. Strong wind resources in the Southwest Power Pool (SPP) lead to exports, despite the CFE deliverability requirement. Higher CFE participation leads to increased energy storage deployment, even though renewables deployment is similar in the 10% and 50% participation cases. Note that the week shown in Figure S9 has very low wind output toward the end of the week. The small magnitudes of hourly CFE demand relative to total load for 10% participation (ranging from 1.3-1.6 GW for most hours in this week) mean that nighttime CFE demand can be met with some energy storage and low wind output, while electrolysis declines to zero. In contrast, the higher CFE demand with 50% participation implies considerably higher energy storage discharge capacity to navigate hourly matching during this challenging week (increasing from 1.6 to 9.6 GW under 10% and 50% participation, respectively).

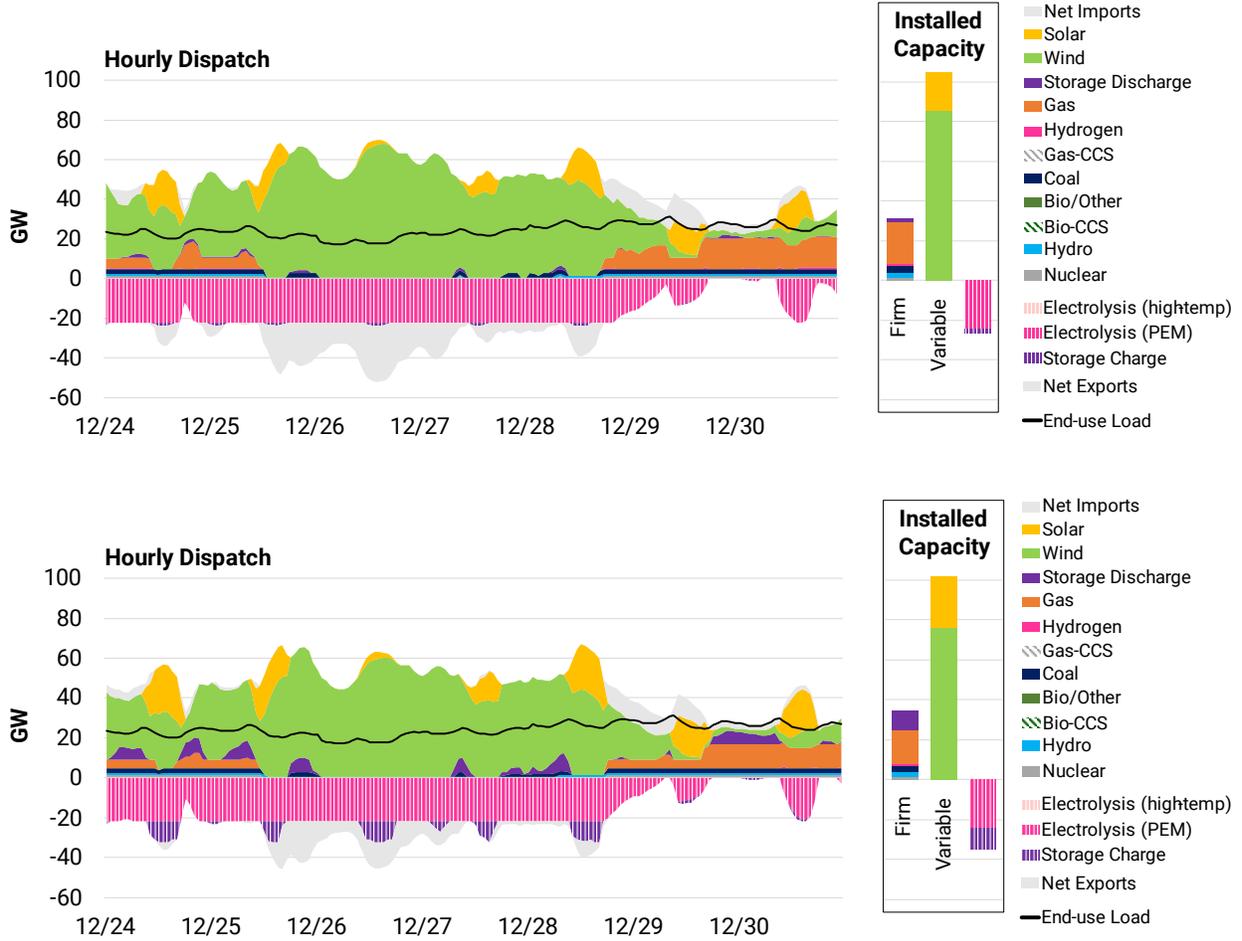

**Figure S9. Hourly dispatch for SPP in 2035 under the three-pillar scenario (cfe_3p) with 10% and 50% participation (top and bottom panels, respectively).** Electrolysis demand, storage charging, and net exports are shown beneath the horizontal axis. Installed capacity is shown on the right panel.

Regional dispatch looks different in the Southeast, which has relatively poor wind resources and only modest solar resources (Figure S10). This solar-dominant system has natural gas resources for firm capacity and generation with a clear diurnal pattern of batteries charging midday and discharging in the evening and nighttime to match CFE demand. With the 50% participation case, there is greater solar, batteries, and nuclear capacity. Installed solar capacity almost doubles, and battery storage nearly doubles. However, gas generation falls faster than gas capacity, as capacity only decreases from 48 GW to 37 GW.

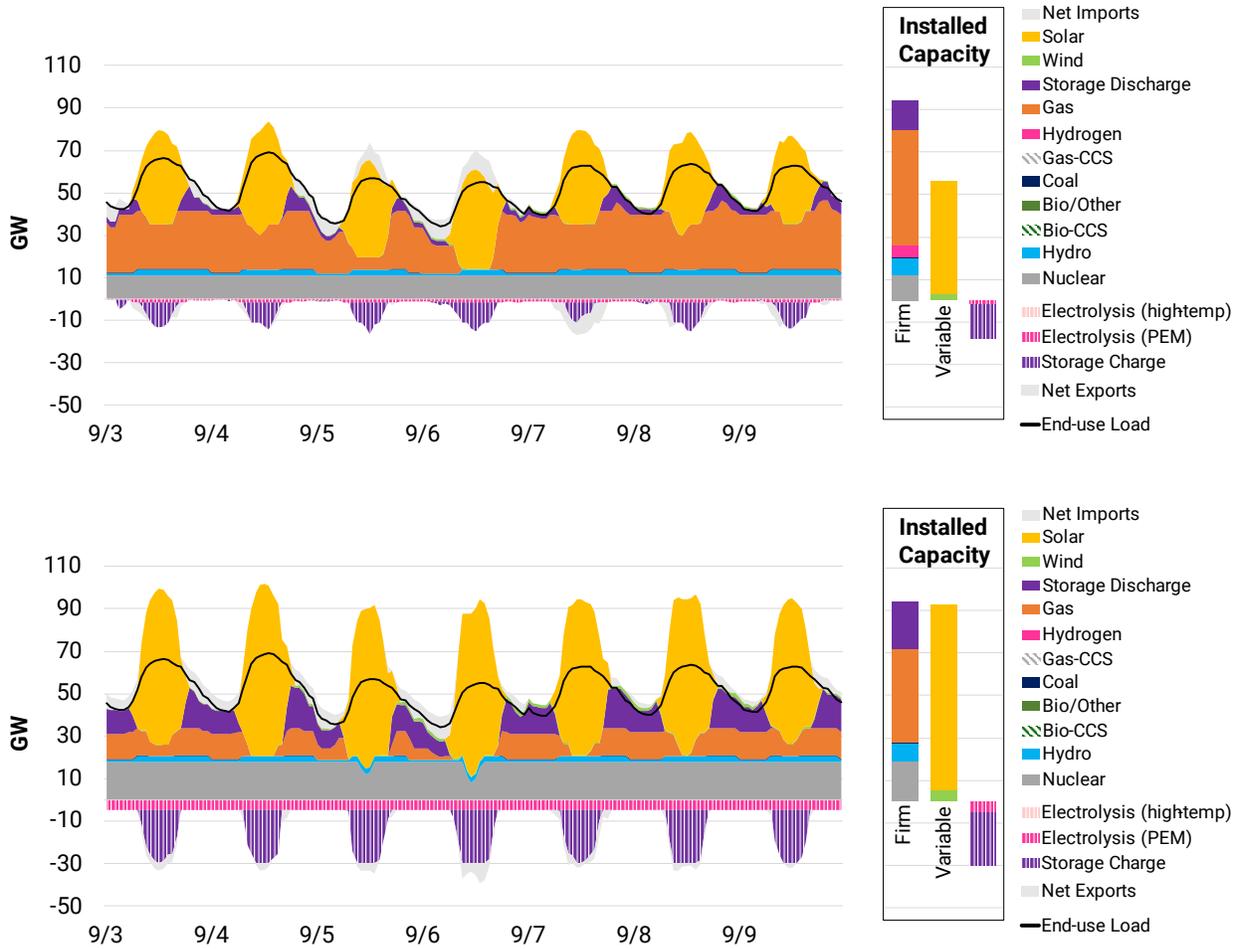

**Figure S10. Hourly dispatch for Southeast in 2035 under the three-pillar scenario (cfe_3p) with 10% and 50% participation (top and bottom panels, respectively).** Electrolysis demand, storage charging, and net exports are shown beneath the horizontal axis. Installed capacity is shown on the right panel.

Regional variation in wholesale electricity prices and EAC prices is shown in Figure S11. Wholesale electricity prices are the consumption-weighted annual average of shadow prices on market-clearing constraints, which do not include EAC prices.

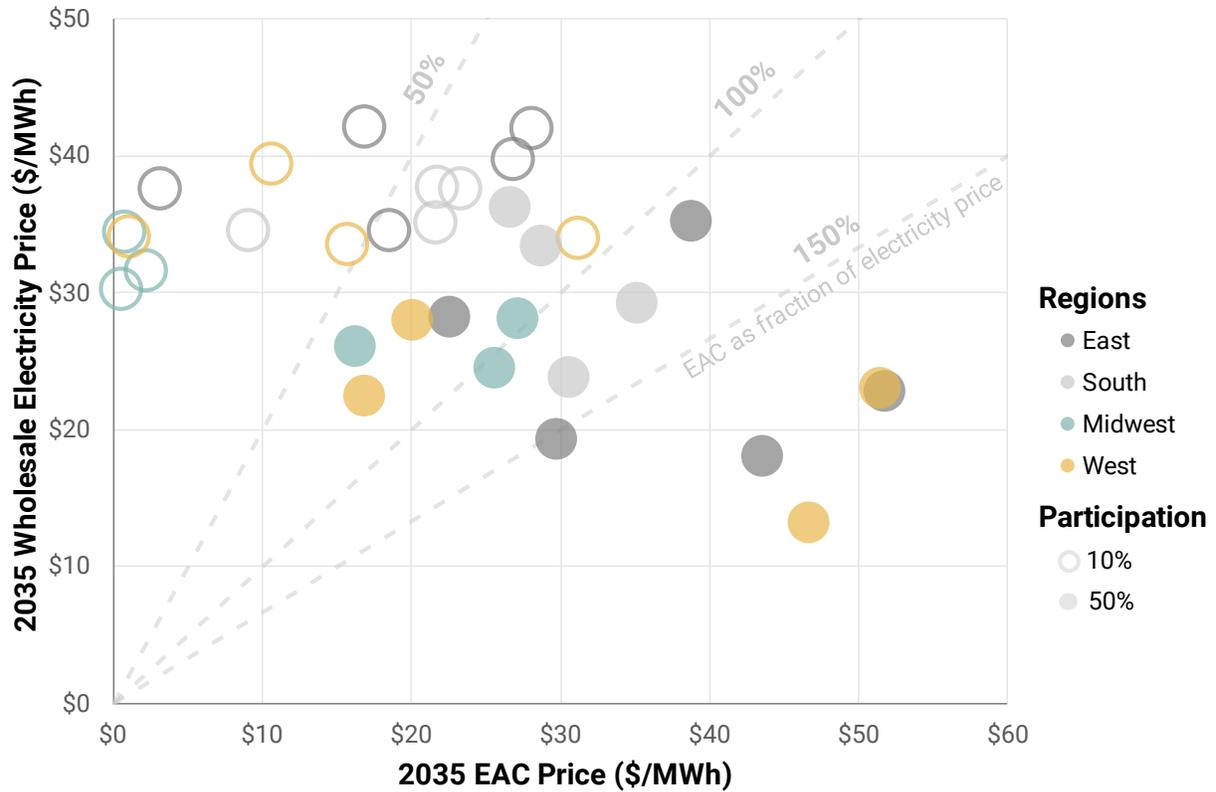

**Figure S11. Regional EAC prices and wholesale electricity prices in 2035.** Reporting regions are shown as colors (Figure S1). 10% participation scenarios are shown as circles, and 50% participation scenarios are shown as dots.

Figure S12 compares regional energy storage deployment, average storage duration, and variable renewables generation across scenarios. Higher energy storage deployment typically occurs for higher solar and wind regions and scenarios. However, there are some conditions with modest energy storage deployment and higher renewables, especially for wind-dominant regions and lower CFE participation. Energy storage deployment and duration typically increases at 50% CFE participation.

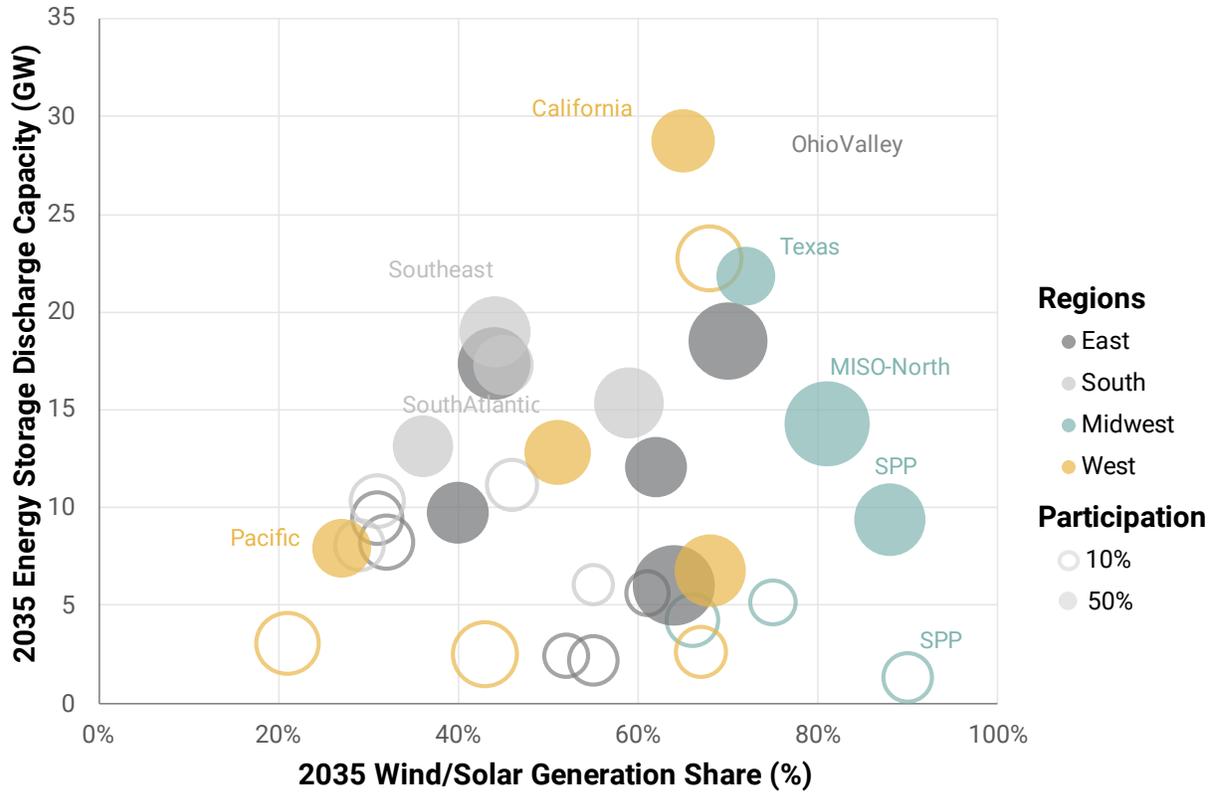

**Figure S12. Regional energy storage deployment and variable renewables generation share in 2035.** Reporting regions are shown as colors (Figure S1). 10% participation scenarios are shown as circles, and 50% participation scenarios are shown as dots. Bubble size is proportional to average regional energy storage duration.

Enforcing qualification criteria has smaller impacts for regions with lower wind and solar resource quality (Figure S13), which have less CFE generation in the reference. In the case with geographical flexibility, the Midwest and West regions are EAC exporters due to their higher quality wind and solar resources. In contrast, the South and East have lower quality wind and solar resources, leading to lower builds in the reference case, which also means that the annual matched case as well as the one with existing resources have larger changes in these regions from the reference. Capacity deployment is also high for the East and South with limited technological options, where incremental capacity to meet the 50% CFE target with three pillars is higher than the Midwest and West.

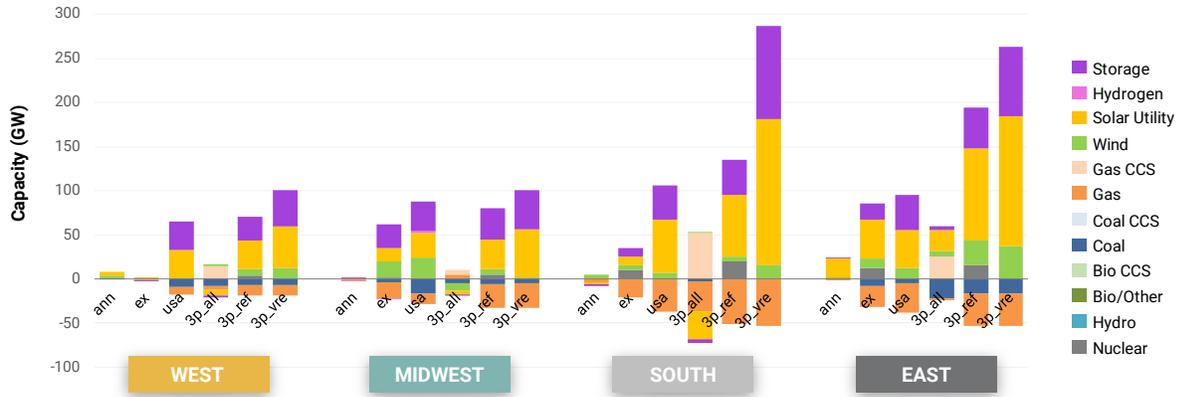

**Figure S13. Regional capacity changes from CFE procurement by technology and scenario in 2035 (relative to the reference without CFE demand).** Changes are shown with 50% C&I CFE demand. CCS = carbon capture and sequestration.

Figure S14 compares generation impacts of CFE demand under different deliverability assumptions. The "cfe_usa" scenario does not enforce the regional deliverability constraint (i.e., allows regional flexibility in meeting the hourly matching and incrementality provisions), "cfe_rr" enforces deliverability with the four reporting regions shown in Figure S1, while "cfe_3p" requires deliverability within the 16 model regions. These scenarios illustrate that using these larger deliverability regions has similar generation (and emissions) outcomes to the more granular deliverability scenario.

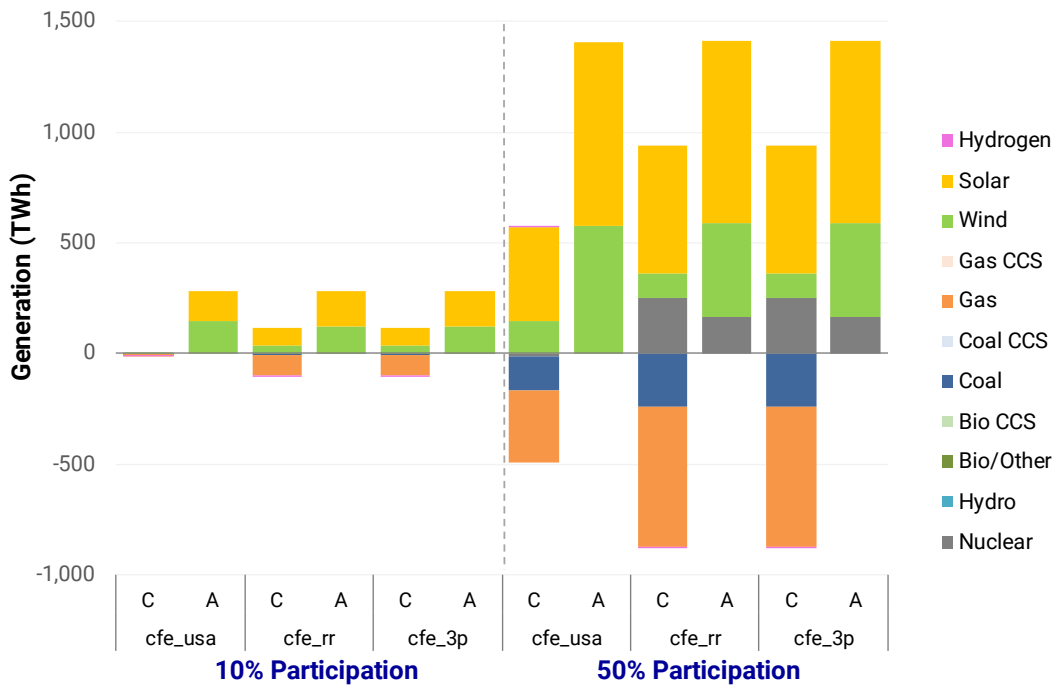

**Figure S14. Generation impacts of CFE demand under alternate deliverability assumptions and participation rates.** Consequential and attributed generation changes by technology and scenario in 2035 (relative to the reference without CFE demand). C = consequential impacts; A = attributed impacts; CCS = carbon capture and sequestration.

*Technology Sensitivities*

Figure S15 compares CFE generation impacts under different technological assumptions. Technology eligibility and availability has the largest impacts with 50% participation. Consequential generation impacts shifts toward CCS-equipped gas generation when the Allam cycle CCS with high $CO_2$ capture rates is available, which is consistent with earlier analysis of EU CFE procurement [8]. When technological portfolios are restricted, wind and solar generation displace some new nuclear and costs increase (Figure 4).

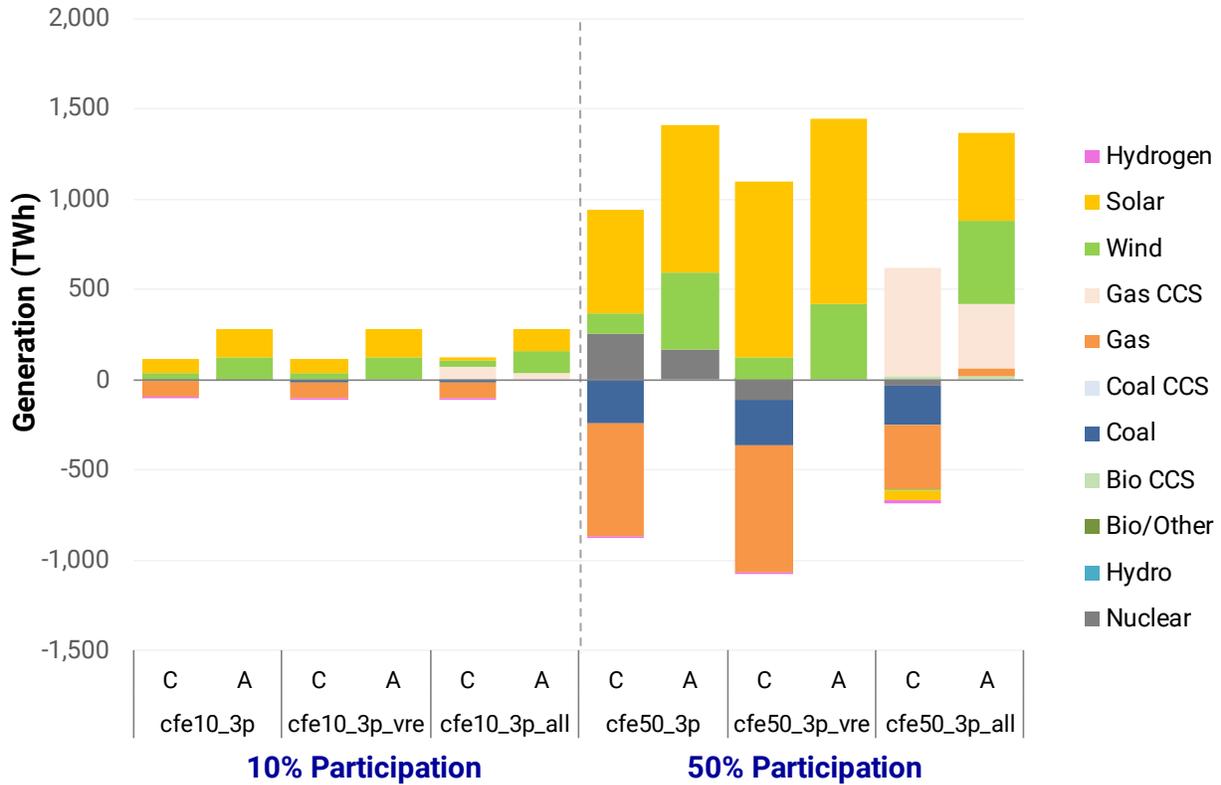

**Figure S15. Generation impacts of CFE demand under alternate technological assumptions and participation rates with three-pillar criteria.** Consequential and attributed generation changes by technology and scenario in 2035 (relative to the reference without CFE demand). C = consequential impacts; A = attributed impacts; CCS = carbon capture and sequestration; VRE = variable renewable energy and batteries only.

*Alternate Weather Years*

Sensitivities to alternate weather years test the robustness of results to inter-annual variability by using alternate weather year data from 1999 through 2019. As shown in Figure S16, different weather years vary in their lengths of wind and solar droughts. The default weather year of 2015 specifically has longer and more frequent wind droughts in several wind-dominant regions.

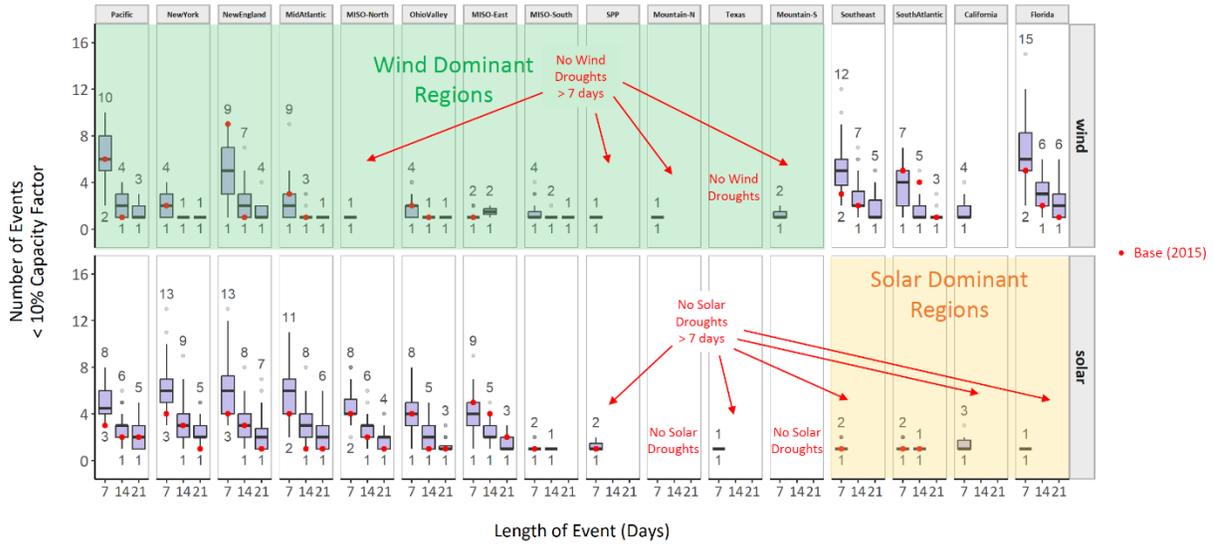

**Figure S16. Number of renewable droughts with less than 10% capacity factor by region and length of event.** Box and whisker plots show values across 1980 through 2019 weather years, where the default 2015 weather year is shown in red. These values are shown for land-based wind (top panel) and utility-scale solar PV (bottom panel). Based on [39].

Although alternate weather years only have a modest impact on national capacity and EAC prices (Figure 5), regional results exhibit greater differences across deployment of specific technologies. Eastern regions have large cross-weather-year differences in solar and energy storage capacity, which exhibits substitutability with nuclear on the margin (Figure S18). Nevertheless, these regions have considerable solar deployment across most weather year scenarios. Note that these weather year sensitivities are conducted for the stringent 50% participation case with three pillars. Scenarios with less stringent CFE procurement targets would likely exhibit more limited differences across weather years.

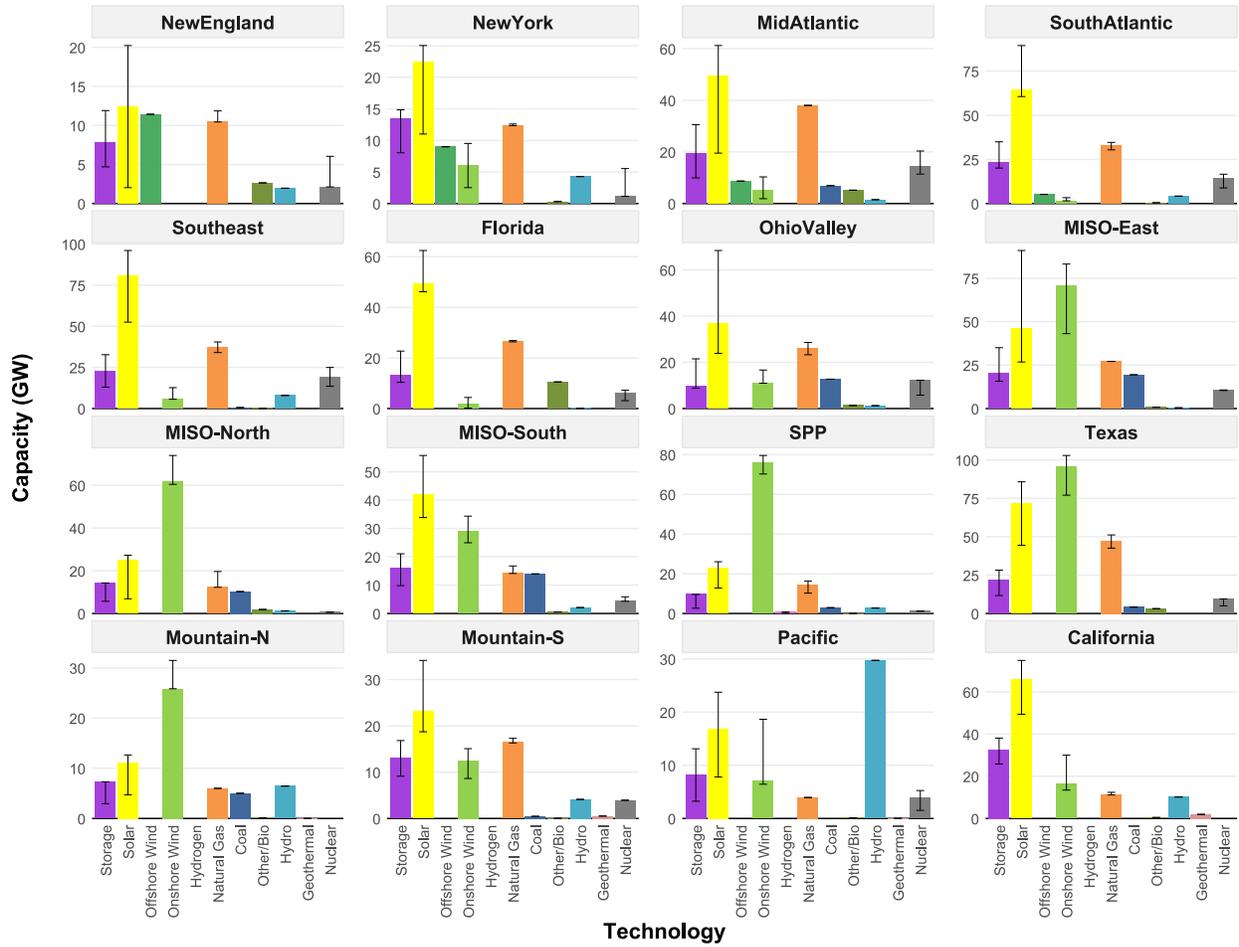

**Figure S17. Impacts of weather years on regional installed capacity by technology for the scenario with three-pillar criteria and 50% participation.** Technology-specific installed capacity is shown for each region. Results for the default 2015 weather year are shown in solid bars, and the differences across 1999 through 2019 weather years are shown as error bars.

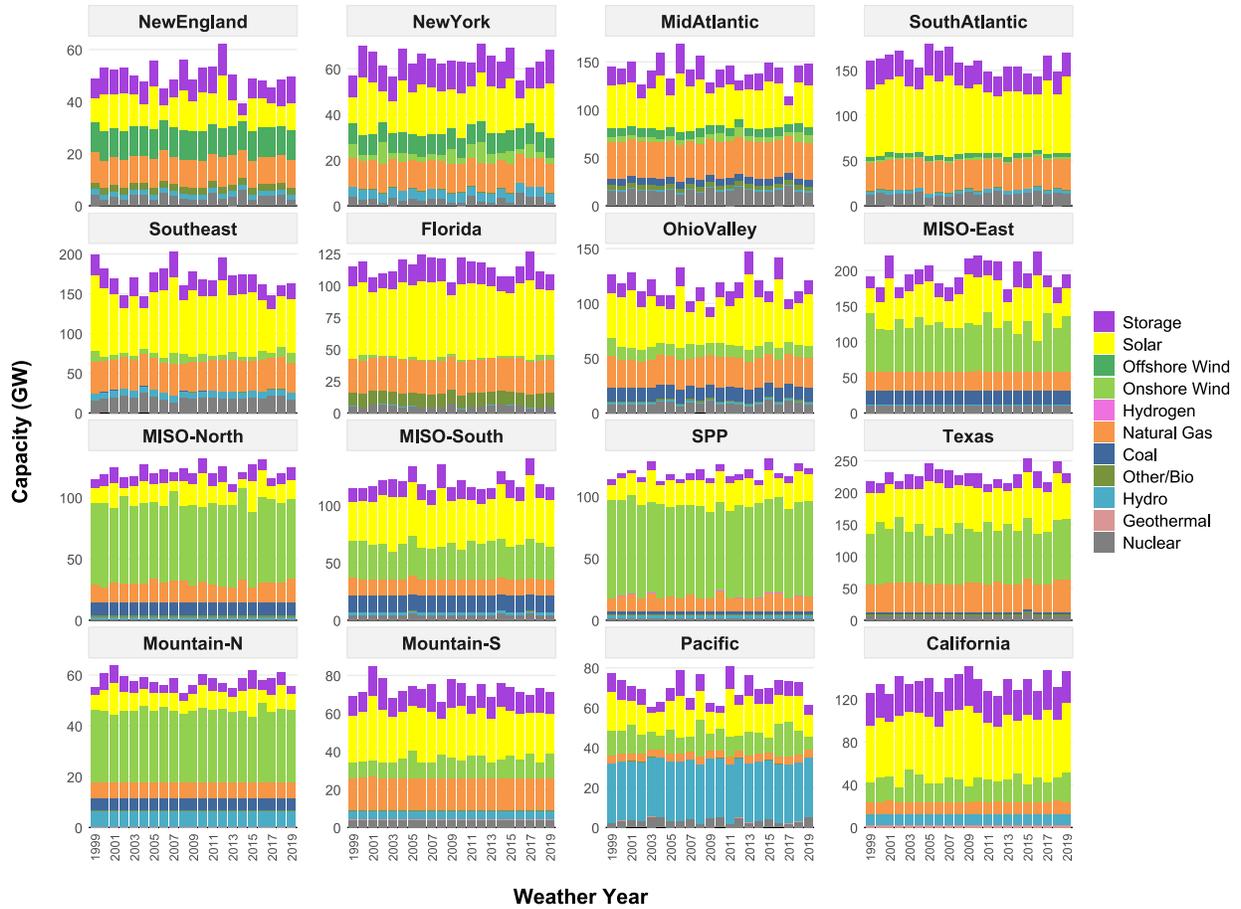

**Figure S18. Impacts of weather years on regional installed capacity by region for the scenario with three-pillar criteria and 50% participation.** Results show 1999 through 2019 weather years.

*Alternate Model Specifications*

Figure S19 compares generation changes from CFE procurement under annual matching ("ann"), hourly matching with flat CFE demand profiles ("3p_flat"), and hourly matching with dynamic CFE profiles from the REGEN end-use model ("3p"). Generation and emissions impacts are similar between the flat and dynamic load profiles. At a national level, matching dynamic profiles leads to slightly greater clean electricity deployment and greater displaced fossil generation, though these effects are relatively small, especially in the 10% participation scenario.

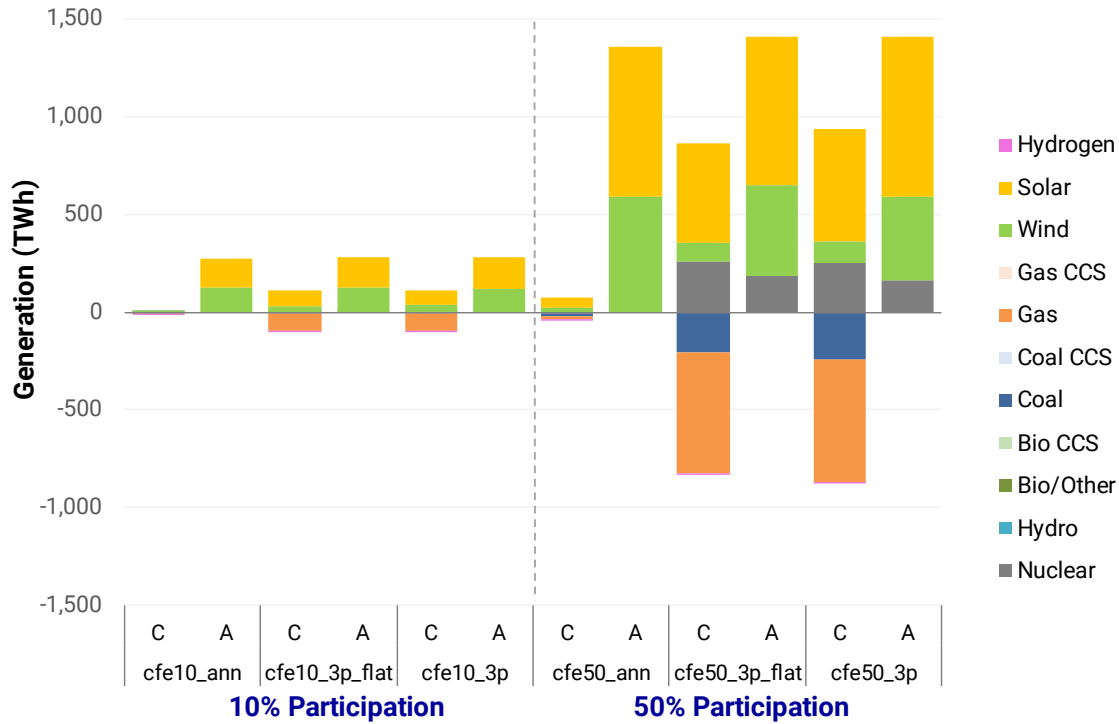

**Figure S19. Generation changes of CFE procurement by technology and scenario in 2035 under alternate load shapes with flat demand and hourly profiles.** Panels show generation changes relative to the reference without CFE demand. C = consequential impacts; A = attributed impacts; CCS = carbon capture and sequestration.

Most results in the paper use the hourly version of REGEN with 8,760 segments for investment and system operations in single future year (labeled "dynamic" in Figure S20). Figure S20 shows results of additional sensitivities that use an intertemporal version of REGEN that optimizes in five-year periods through 2050 with 120 intra-annual periods and reduced-form chronology. The hourly model generally has greater solar and energy storage deployment, which lowers wind generation.

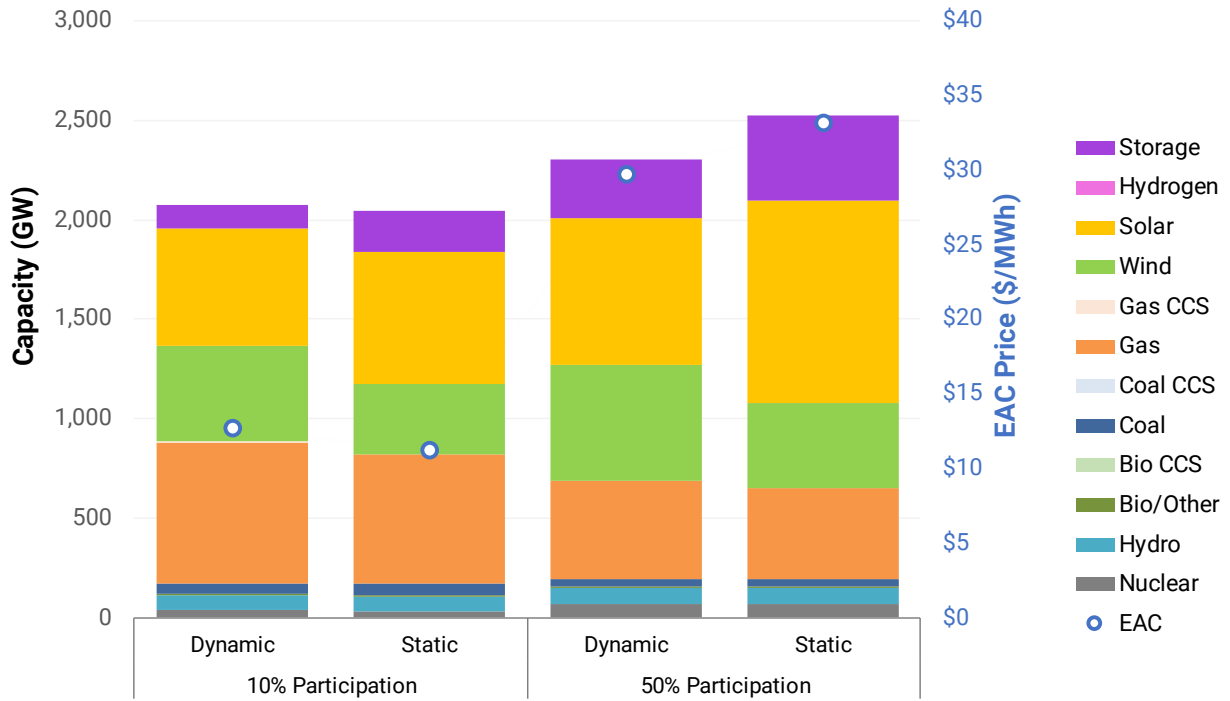

**Figure S20. Installed capacity in 2035 under different model temporal resolutions and alternate CFE participation rates.** Scenarios with three-pillar criteria are shown. EAC prices are shown on the secondary axis. CCS = carbon capture and sequestration.

*Policy Sensitivities*

The policy sensitivities in Figure S21 show how the background policy environment alters generation and emissions implications of CFE procurement. Differences are largest with less stringent climate policies and incentives, which brings less low-emitting generation under baseline conditions.

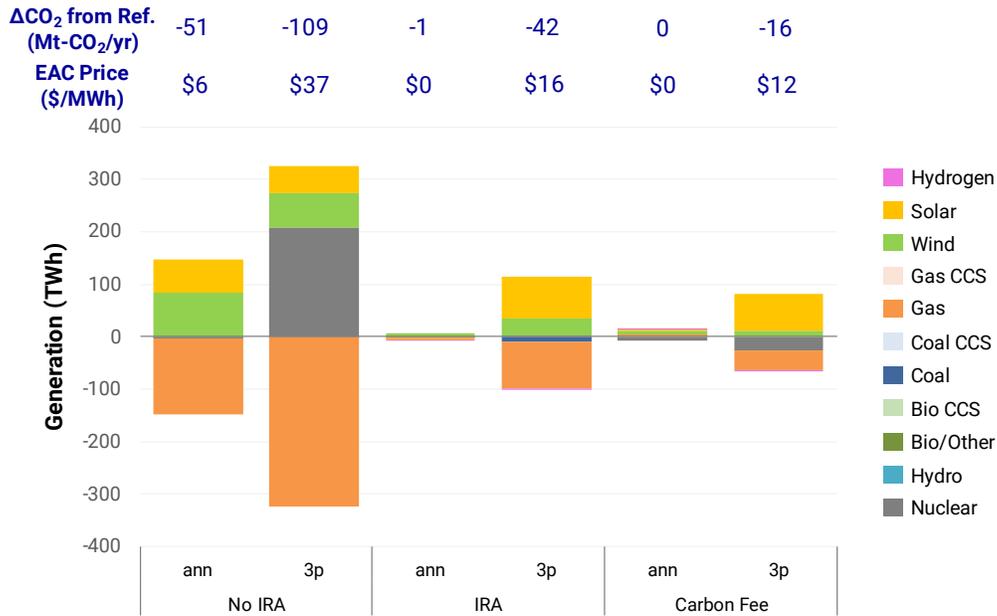

**Figure S21. Generation changes of CFE procurement by technology and scenario in 2035 under alternative policy environments.** Panels show generation changes relative to the reference without CFE demand under annual (ann) and hourly matching (3p) assuming 10% participation rate. (B) Sensitivities to load shapes with flat demand and hourly profiles. CCS = carbon capture and sequestration; IRA = Inflation Reduction Act.

**Supplementary Note 3: List of Abbreviations**

| | |
|---|---|
| 3P | Scenario where all three pillars for CFE procurement are enforced (temporal matching, incrementality, and deliverability) |
| 45Q | Inflation Reduction Act tax credit for captured $CO_2$ |
| 45V | Inflation Reduction Act clean hydrogen production tax credit |
| 45Y | Inflation Reduction Act clean electricity production tax credit |
| 48E | Inflation Reduction Act clean electricity investment tax credit |
| A | attributed |
| C | consequential |
| C&I | commercial and industrial sectors |
| CCS | carbon capture and storage |
| CDR | carbon dioxide removal |
| CES | clean electricity standard |
| CFE | carbon-free electricity |
| $CO_2$ | carbon dioxide |
| EAC | energy attribute certificate |
| EPRI | Electric Power Research Institute |
| GW | gigawatt |
| $H_2$ | hydrogen |
| IRA | Inflation Reduction Act of 2022 |
| MISO | Midcontinent Independent System Operator |
| Mt | million metric tonnes |
| MWh | megawatt-hour |
| NGCC | natural gas combined cycle |
| PV | photovoltaic |
| RPS | renewable portfolio standard |
| SPP | Southwest Power Pool |
| US-REGEN | U.S. Regional Economy, Greenhouse Gas, and Energy |
| USD | United States dollar |